\DocumentMetadata{}
\documentclass[sigconf, 9pt]{acmart}
\usepackage{textcomp}
\usepackage{enumitem}
\usepackage{hyperref}
\usepackage{hyperxmp} 
\usepackage{hyphenat}

\usepackage{threeparttable}
\usepackage{booktabs}
\usepackage{siunitx}
\usepackage{multirow}
\usepackage{graphicx}
\usepackage{caption}
\usepackage[normalem]{ulem}
\usepackage{amsmath}

\AtBeginDocument{%
  \providecommand\BibTeX{{%
    \normalfont B\kern-0.5em{\scshape i\kern-0.25em b}\kern-0.8em\TeX}}}
\usepackage{multirow}

\copyrightyear{2026}
\acmYear{2026}
\setcopyright{cc}
\setcctype{by}
\acmConference[E-Energy '26]{The 17th ACM International Conference on Future and Sustainable Energy Systems}{June 22--25, 2026}{Banff, AB, Canada}
\acmBooktitle{The 17th ACM International Conference on Future and Sustainable Energy Systems (E-Energy '26), June 22--25, 2026, Banff, AB, Canada}
\acmDOI{10.1145/3744255.3811711}
\acmISBN{979-8-4007-2011-6/2026/06}

\begin{document}


\title{Semantic Technologies in Practical Demand Response: An Information Requirement-based Roadmap}

\author{Ozan Baris Mulayim}
\email{omulayim@andrew.cmu.edu}
\authornote{Corresponding Author}
\affiliation{%
  \institution{Carnegie Mellon University}
  \streetaddress{}
  \city{Pittsburgh}
  \state{PA}
  \country{United States}
  \postcode{15217}
 } 

\author{Anand Krishnan Prakash}
\email{akprakash@lbl.gov}
\affiliation{%
  \institution{Lawrence Berkeley National Laboratory / Carnegie Mellon University}
  \streetaddress{}
  \city{Berkeley}
  \state{CA}
  \country{United States}
  \postcode{94720}
 } 

\author{Yuvraj Agarwal}
\email{yuvraj@cs.cmu.edu}
\affiliation{%
  \institution{Carnegie Mellon University}
  \streetaddress{5000 Forbes Avenue}
  \city{Pittsburgh}
  \state{PA}
  \country{United States}
  \postcode{15213}
 }

\author{Mario Berg{\'e}s}
\email{marioberges@cmu.edu}
\authornote{Mario Berg{\'e}s holds concurrent appointments as a Professor of Civil and Environmental Engineering at Carnegie Mellon University and as an Amazon Scholar. This paper describes work at Carnegie Mellon University and is not associated with Amazon. }
\affiliation{%
  \institution{Carnegie Mellon University}
  \streetaddress{}
  \city{Pittsburgh}
  \state{PA}
  \country{United States}
  \postcode{15217}
}

\author{Marco Pritoni}
\email{mpritoni@lbl.gov}
\affiliation{%
  \institution{Lawrence Berkeley National Laboratory}
  \streetaddress{}
  \city{Berkeley}
  \state{CA}
  \country{United States}
  \postcode{94720}
 } 

\author{Derek Supple}
\email{derek.supple@jci.com}
\affiliation{%
  \institution{Johnson Controls}
  \streetaddress{507 E Michigan St}
  \city{Milwaukee}
  \state{WI}
  \country{United States}
  \postcode{53202}
 }

\author{Steve Schaefer}
\email{stephen.ashley.schaefer-ext@jci.com}
\affiliation{%
  \institution{Johnson Controls}
  \streetaddress{507 E Michigan St}
  \city{Milwaukee}
  \state{WI}
  \country{United States}
  \postcode{53202}
 }

\author{Mitali Shah}
\email{mitali.shah@itron.com}
\affiliation{%
  \institution{ITRON}
  \streetaddress{2111 N Molter Road}
  \city{Liberty Lake}
  \state{WA}
  \country{United States}
  \postcode{99019}
 }


\renewcommand{\shortauthors}{Mulayim et al.}
\begin{abstract}

The transition to a modern and efficient future grid relies on the seamless coordination of distributed energy resources and applications such as Demand Response (DR). While this transformation enables greater sustainability, it inevitably increases grid complexity and decentralization, requiring the effective coordination of millions of hardware assets and software agents. Realizing this vision demands advances in interoperability to ensure these heterogeneous systems can communicate without prohibitive customization costs. Semantic interoperability aims to address this by leveraging ontologies to guarantee the unambiguous interpretation of exchanged data. However, current semantic ontologies in the commercial building and DR domains face two critical limitations. First, existing ontologies are often developed without a formal framework that reflects real-world DR requirements. Second, proposals for integrating general (e.g., Brick) and DR-specific ontologies (e.g., EFOnt) remain mostly conceptual, lacking formalization or empirical validation.

In this paper, we begin to address these gaps by applying a formal ontology evaluation/development approach to define the information requirements (IRs) necessary for semantic interoperability, focusing on incentive-based DR programs for commercial buildings in the United States as a starting point. We identify the IRs associated with each stage of the incentive-based DR. Using these IRs, we evaluate how well existing ontologies, specifically Brick, DELTA, EFOnt, and CIM support the operational needs of DR participation. Our findings reveal substantial gaps between current ontologies and practical DR requirements. Based on our evaluation, we propose a roadmap of necessary extensions and integrations for these ontologies. This work ultimately aims to enhance the interoperability of today's and future smart grid, thereby facilitating scalable integration of DR systems into the grid's complex operational framework.

\end{abstract}

\ccsdesc[500]{Hardware~Smart grid}
\ccsdesc[500]{Information systems~Ontologies}
\keywords{Semantic Ontology, Demand Response, Information Requirements}


\maketitle
\section{Introduction}
\label{sec:introduction}

The grid of the future will be significantly more complex and decentralized than today's grid due to the increasing penetration of distributed energy resources (DERs) \cite{shaukat2023decentralized}. 
Rather than investing in additional generation capacity or distribution network upgrades, grid operators can control flexible assets to provide grid services through demand response (DR) programs \cite{poudineh2014distributed}.
However, realizing this flexibility at scale presents a significant challenge. 
Coordinating millions of these flexible assets requires software platforms capable of reliably responding to grid signals across multiple stakeholders and interfaces \cite{howell2017towards}. 
Additionally, the software solutions being developed for each stakeholder and their assets (e.g., building operators, DR service providers and aggregators) are independent of the others and are not interoperable; each having their own business rules and information requirements (IRs). 
Although many of these requirements are individually known within each stakeholder community (for example, grid operators understand program rules and Curtailment Service Providers (CSPs) understand aggregation logic), a unified, formally-grounded view that spans the complete DR lifecycle is lacking.
These software solutions include, for example, software interacting with building management systems to allow for assets to respond to grid service requests or to estimate load flexibility and submit bids to participate in DR markets.
As a result, these siloed DR solutions ecosystem requires expensive, ad-hoc engineering efforts that are difficult to replicate and scale.
This ultimately hinders the participation of devices in DR programs and limits their potential to provide grid services. 

Semantic interoperability or the ability of systems to exchange information with shared, unambiguous meaning has successfully addressed similar integration challenges in domains such as healthcare, manufacturing, and building automation (e.g.,~\cite{balaji2016brick}). 
By establishing common data models and standardized vocabularies, semantic interoperability enables diverse systems to communicate effectively without requiring custom integration for each connection. 
Such approaches have the potential to lower the high cost of entry for DR participation \cite{coe_demanding_2010} and reduce the expense of achieving automation by enabling application portability across assets \cite{bergmann2020semantic}. 
While interoperability challenges in other industries have been addressed through standardization efforts and semantic modeling technologies, similar challenges in the grid services sector remain largely unsolved \cite{Kim2017Toward, LBNLGEB2023, NISTSP1108}. 
Commercial platforms supporting DR programs through proprietary mechanisms can also benefit from such standardization efforts as it will reduce implementation costs, lower barriers to entry and expand their customer base.
Hence, the technologies required to bring about this future grid involve not only hardware advancements but also a leap in interoperable software solutions grounded in semantic standards that can bridge the fragmented DR ecosystem.

Significant efforts have been made to advance semantic technologies for commercial buildings and DR applications \cite{hoare2022linked, tomavsevic2015ontology, de2025semantics, kuvcera2018semantic}. Nevertheless, existing ontologies have been developed with distinct objectives, limiting their ability to fully support the diverse IRs and varying business rules found across different DR stakeholders \cite{pritoni_metadata_2021, coe_demanding_2010}. For semantic models to be truly effective, their underlying ontologies must align with and comprehensively cover the data requirements of the software applications they are designed to support \cite{beydoun2024tailoring}. Building ontologies such as Brick \cite{balaji2016brick} and Haystack \cite{ProjectHaystack} have been designed to provide interoperability for commercial buildings and software applications with these buildings, but they lack the specificity required for DR-specific use cases. Conversely, DR-focused ontologies such as EFOnt \cite{li_semantic_2022} and DELTA \cite{fernandez2021supporting} emphasize energy flexibility but do not comprehensively capture the operational and data requirements of incentive-based DR \footnote{We define incentive-based DR later in the manuscript.}. 
Focusing more on the grid, the Common Information Model (CIM)~\cite{uslar2012common} captures market interaction and program participation requirements well, but lacks coverage of DR-specific requirements.

A potential solution to these challenges lies in combining a building-centric ontology with a DR-specific ontology and ensuring that they jointly cover the IRs of automated DR applications. While this idea was conceptually introduced in IEA EBC Annex 81 \cite{johra2023iea, IEA_Annex81}, it has neither been formalized nor empirically validated against the practical information needs of DR programs. Moreover, existing ontologies have not been developed using a formal framework that prioritizes practical implementation constraints. For instance, EFOnt was designed based on data requirements for EnergyPlus simulations, which differ fundamentally from the operational requirements of physical commercial buildings. Similarly, recent evaluations of Brick’s coverage for Model Predictive Control (MPC) applications found significant gaps in its ability to support key control requirements for DR \cite{prakash2024ontologies}. Consequently, without a systematic framework that ensures ontologies satisfy practical decision-making requirements, simple integration will likely fall short of enabling seamless semantic interoperability.

Recognizing this gap, our research formalizes this ontology integration approach by defining the IRs necessary for semantic interoperability in \textit{incentive-based DR for commercial buildings}. Originating from the systems and software engineering domains, IRs define the data elements, dependencies, and constraints that enable structured information exchange between systems \cite{lana2021data, sommerville2009deriving, ma2024state}. We specifically consider the IRs needed for decision-making during each stage of the business process of incentive-based DR as defined by \cite{NAESBSGTF2010, coe_demanding_2010}. 

We explicitly delimit the scope of this analysis to DR programs administered by United States of America (USA)-based Independent System Operators (ISOs), targeting commercial buildings as the primary actors. We acknowledge that the specific regulatory constraints and market definitions identified herein may not directly apply to other geographies or building sectors (e.g., residential). However, the primary goal of this work is to provide a comprehensive analysis within this reduced scope while demonstrating a rigorous method for ontology evaluation. This methodological framework is designed to be transferable, allowing future research to apply the same principles to other contexts and wider scopes. 

\begin{figure}[h!]
\centering
\vspace{-3ex}
\includegraphics[width=1\columnwidth]{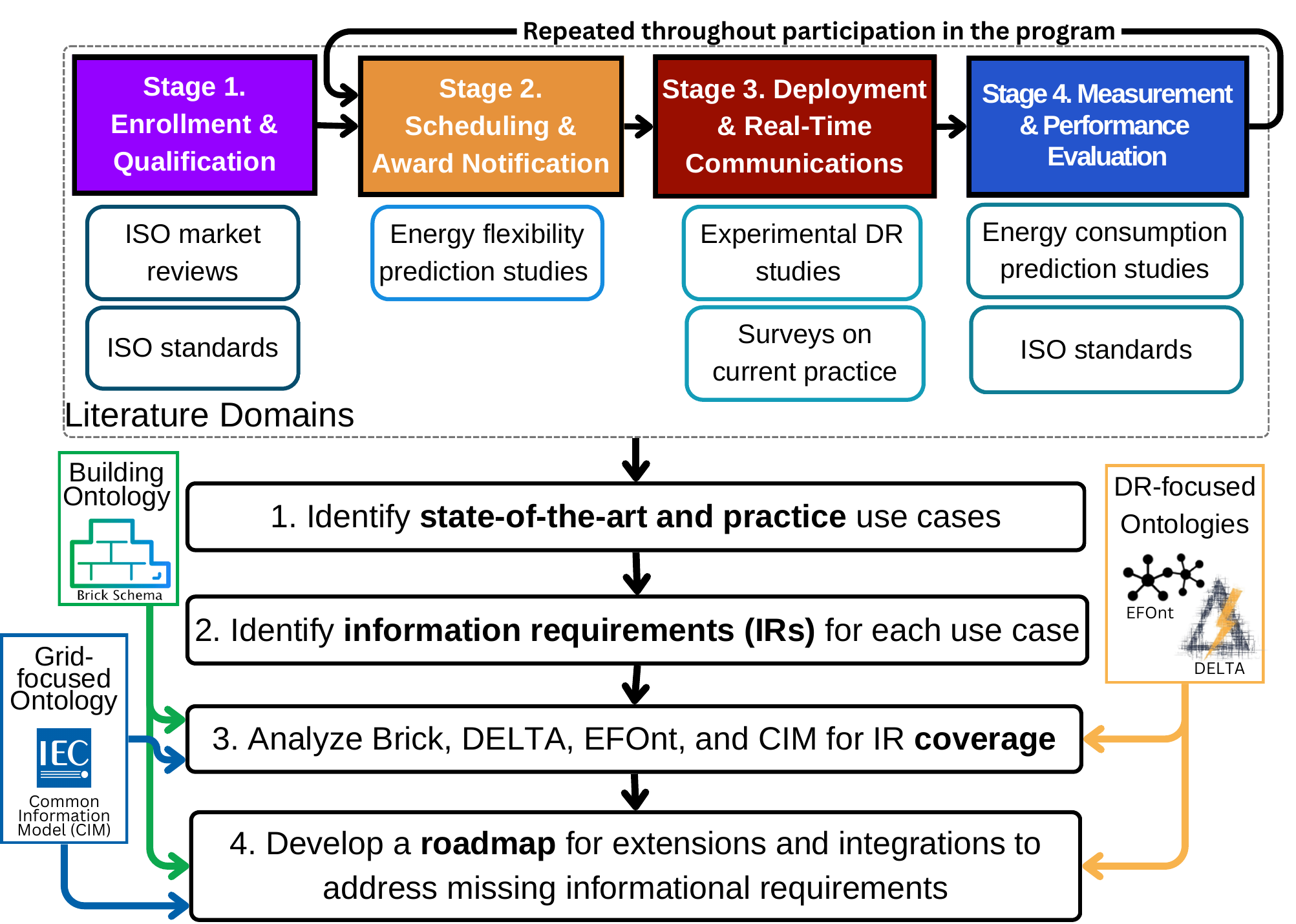}
\caption{Workflow for reviewing, analyzing and extending the coverage of existing ontologies for practical DR applications.}
\label{fig:questions}
\end{figure}

We start by identifying the existing information exchanges mandated by current DR program manuals and guidelines in the USA. However, relying solely on today's documentation risks immediate obsolescence, as the grid transitions toward sophisticated agents utilizing predictive control and data-driven flexibility forecasting. To account for this evolution, we expand our analysis to include data requirements derived from experimental algorithms and advanced methodologies found in the literature, rather than limiting our scope to current compliance standards. Our primary goal is to establish a robust foundation for interoperability standards that supports both present-day operations and future automation\footnote{Note that this list is not expected to be exhaustive; much like software, ontologies require continuous updates to remain relevant \cite{pritoni_metadata_2021}.}. We then apply these IRs to evaluate the coverage of the commercial building ontology Brick alongside the DR-focused ontologies DELTA and EFOnt and the grid-focused CIM. While the combination of DELTA and EFOnt might, in principle, satisfy all IRs, our findings reveal significant gaps. Consequently, we analyze how to represent the unsatisfied IRs with extensions to their infrastructure. Figure \ref{fig:questions} provides an overview of our analysis. This paper employs a formal ontology-development framework based on IRs and proposes a future roadmap of extensions to the current ontology infrastructure. 
We make the following contributions in this paper: 

\begin{itemize}
\item For each stage of DR, we review the literature and existing US ISO programs such as those from PJM \cite{PJM2024Manual11, PJM2024DemandResponse} to identify the specific IRs needed to participate in that stage (as defined by the North American Energy Standards Board (NAESB) \cite{NAESBSGTF2010, coe_demanding_2010}).
\item We evaluate the coverage of four ontologies (Brick, EFOnt, DELTA and CIM) across the identified IRs.
\item We outline a roadmap for future ontological development in the building space to ensure complete coverage of the identified IRs, moving us closer to the vision of Grid-Interactive Efficient Buildings.
\end{itemize}

\textbf{Vision:} By developing these IRs and improving the underlying ontologies, we envision a future ecosystem grounded in comprehensive semantic models of both DR processes and commercial buildings. This standardization enables entities, such as aggregators, to seamlessly scale their portfolios by adding new buildings and binding them to uniform DR applications. Ultimately, this allows for the coordination and control of diverse assets using a single machine-readable language, replacing bespoke integrations with universal interoperability.

\section{Background}
\label{sec:background}

This section establishes the foundational context for understanding interoperability challenges in DR. We first define the ecosystem of stakeholders and information exchanges governing DR participation (Section \ref{sec:interactions}), followed by a characterization of the operational requirements for incentive-based DR programs (Section \ref{sec:character_background}). Finally, we review the current landscape of semantic ontologies, identifying gaps in both broad building-centric models and specialized DR frameworks (Section \ref{sec:ontologies_background}).

\subsection{Information Exchange in Demand Response Ecosystems} \label{sec:interactions}

In the United States, the administration of DR programs depends heavily on the regional market structure. While Independent System Operators (ISOs) and Regional Transmission Organizations (RTOs) manage wholesale markets that cover the majority of the load, they are not the sole administrators. Utilities also play a central role, directly administering retail DR programs within ISO regions and serving as the primary program administrators in vertically integrated regions without an ISO/RTO. Consequently, participation pathways vary: retail DR services are typically procured and managed directly by utilities, whereas wholesale DR services are procured in energy markets through intermediaries known as aggregators or CSPs.

\begin{figure}[b] \centering
\includegraphics[width=0.8\columnwidth]{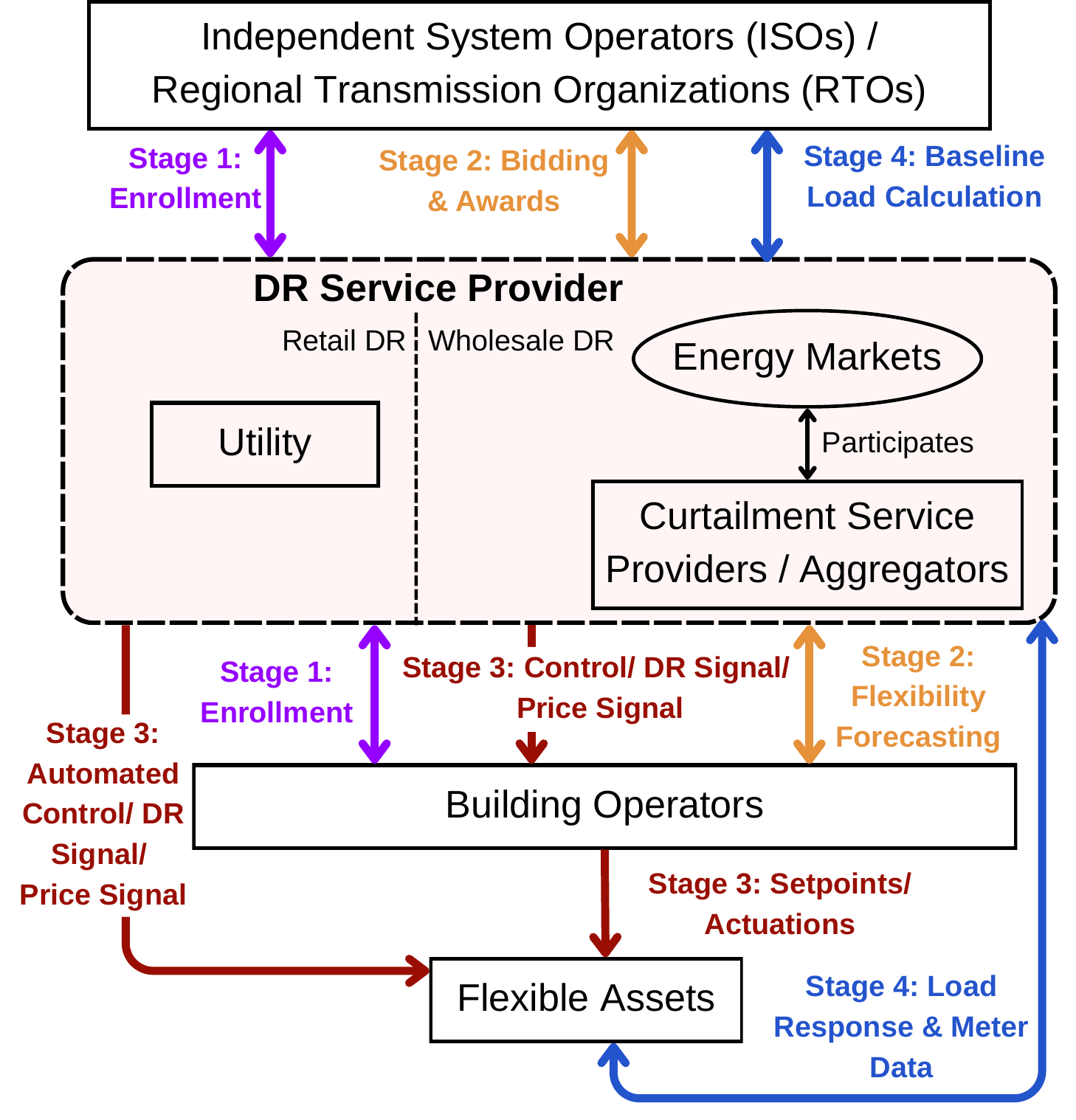}
\caption{The information exchange DR ecosystem between US grid entities and building assets. The arrows indicate specific substeps required at each of the four stages of DR participation.} \label{fig:interaction_flow}
\end{figure}

Figure \ref{fig:interaction_flow} illustrates the coordination required across these primary stakeholders. The diagram distinguishes between the Retail and Wholesale pathways (enclosed in the dashed box) and maps the information flow down to the building operators and physical assets. Crucially, the color-coded arrows represent the specific applications and substeps required within each of the four DR stages.

\textbf{Grid Operators (ISO/RTO):} As the entities responsible for maintaining grid reliability in energy markets, ISOs and RTOs function as the market makers for DR. They define the rigorous enrollment criteria (Stage 1) and clear the market to issue commitments (or awards) based on bids (Stage 2). Crucially, these day-ahead commitments are distinct from the actual dispatch instructions issued during the operating window to balance supply and demand (Stage 3). Finally, they facilitate financial settlement (Stage 4) based on verified performance.

\textbf{Curtailment Service Providers (CSPs) / Aggregators:} Most commercial buildings cannot meet the minimum resource size to participate directly in DR programs. The CSP bridges this gap by acting as an intermediary. For example, during the scheduling and bidding stage (Stage 2), the CSP aggregates flexibility forecasts from multiple buildings to construct a compliant market bid. They function as a complexity buffer, translating the ISO’s economic dispatch instructions into site-specific control signals that building operators can act upon in Stage 3.

\textbf{Utilities:} The utility's role varies significantly depending on the market structure. In vertically integrated regions without an ISO/RTO, the utility functions as the sole grid operator, responsible for balancing supply and demand directly (Stage 3). In wholesale markets, they primarily support the measurement and performance evaluation stage (Stage 4) by supplying the revenue-grade interval meter data necessary for calculating the customer baseline load (CBL). However, in Retail DR programs, the utility effectively functions as the CSP. In this capacity, they manage participation of individual customers, performing the forecasting and dispatch functions that a CSP would otherwise handle in the wholesale market.

\textbf{Building Operators (BO):} At the edge of the grid, building operators serve as the interface between the economic logic of the market and the control logic of the building. 
When participating in DR programs is enabled through manual interventions, this role requires translating external grid service signals into internal system commands and executing specific setpoints and actuations to modulate flexible assets. Alternatively, for automated DR programs, operators must configure their systems to receive and automatically execute control signals sent directly by utilities or CSPs. In both scenarios, they must protect the comfort and productivity of building occupants, ensuring that any market participation respects internal operational constraints.

In this work, we anchor our analysis on the interfaces at the two ends of this ecosystem: the codified program standards defined by ISOs and the operational strategies of building assets described in the literature. While our review encompasses studies on market structures, we do not explicitly model the internal data architectures of CSPs and utilities. As commercial entities, their specific implementations are often proprietary and highly variable, lacking the standardized public documentation available for ISO markets or the physics-based consistency of building control. However, regardless of the aggregation logic employed, the fundamental information exchange is bounded by the ISO market's requirements and the building's physical capabilities. By focusing on these primary constraints, we aim to derive a robust set of IRs that underpin the entire value chain.
\vspace{-3ex}

\subsection{Characterization of DR programs}\label{sec:character_background}
DR programs, designed to encourage changes in electricity consumption patterns in response to grid conditions or price signals, vary significantly in their specifics from one ISO to another \cite{faria_overview_2015, macdonald_demand_2023, helman_demand_2021}. An essential aspect here is the industry's use of the term ``real-time.'' Typically, ``real-time'' refers to notifications issued a few hours in advance — a timeframe more accurately described as intra-day. To clarify this terminology within our discussions, we use ``instantaneous'' to denote truly real-time responses. 

DR programs can be primarily classified based on the trigger for participation: price-based or incentive-based \cite{USDOE2006}. Price-based DR programs leverage time-variable electricity pricing (e.g., Time-of-Use rates or Critical Peak Pricing) to encourage voluntary demand reduction during high-price periods; for example, a commercial building might reduce its HVAC load when electricity prices are exceptionally high to lower its energy bill. In contrast, incentive-based DR programs offer participants direct financial incentives, such as bill credits or payments, for verified load reductions during specific DR events when dispatched by the utility or ISO. Participation in such programs often involves meeting minimum resource size limitations—a key constraint verified during the enrollment phase (Stage 1)—sometimes necessitating aggregation through a CSP. Although our current scope is primarily incentive-based DR, the principles discussed could be extended to price-based DR in future work.

Within the broad category of incentive-based DR, programs are further distinguished by the service function they provide to the grid. These services are commonly categorized as energy services, capacity services, and ancillary services \cite{cappers_assessment_2013}, with the latter encompassing regulation and reserves, drawing from established definitions like those by the NAESB \cite{NAESB2008}. A detailed characterization of these services, including their operational definitions and timing parameters, is provided in Appendix \ref{sec:appendix_services}.

\subsection{Semantic Ontologies for Commercial Buildings and Demand Response}\label{sec:ontologies_background}

The field of semantic ontologies in commercial buildings has seen a diverse range of offerings tailored to varying objectives and levels of specificity. These ontologies seek to formalize domain knowledge representation and enable the integration of heterogeneous data in a structured, machine-readable format, ensuring consistency across applications \cite{hartmann2020advanced}. Previous work has typically concentrated on specialized topics within the building sector, including Indoor Environmental Quality \cite{donkers2022semantic}, Fault Detection \cite{benndorf2018energy}, Building Information Modeling (BIM) \cite{venugopal_ontology-based_2015, sacks2022toward, shi2023ontology}, among others. A systematic review by \cite{luo2021overview} highlighted the diversity of available tools: ranging from dictionaries to ontologies and platforms, underscoring the fragmented nature of current research. Pritoni et al. \cite{pritoni_metadata_2021} expanded this analysis by examining real-world use cases, revealing that many ontologies lack active community engagement and struggle to support the evolving schema diversity required in building applications. Comparative evaluations of Brick 1.0 and Project Haystack 3.0 indicate Brick's superior completeness and expressiveness for Smart Building applications \cite{quinn_case_2021}. However, recent research assessing Brick’s coverage for MPC use cases found that it fails to meet key requirements \cite{prakash2024ontologies}, exposing gaps that limit its practical application. These challenges have prompted efforts to classify and better understand the roles of different types of ontologies.  

Building on this landscape, we categorize the relevant models based on their primary domain focus. We identify widely adopted standards like Brick \cite{balaji2016brick} and Haystack \cite{ProjectHaystack} as building-centric ontologies, serving as the broad foundation for representing commercial building data. In contrast, we classify models like DELTA \cite{fernandez2021supporting} and EFOnt \cite{li_semantic_2022} as DR-specific ontologies and CIM~\cite{uslar2012common} as grid-specific, as they focus explicitly on the application layer of DR and coordination of the grid power systems and energy markets, respectively. 

Despite the growing number of proposed ontologies, few achieve widespread acceptance. The absence of a centralized repository complicates their discovery \cite{summit2016ontologies}, and Pritoni et al.~\cite{pritoni_metadata_2021} found that when examining 70 ontologies listed on a website, many were inaccessible due to expired links, inadequate documentation, or language barriers, highlighting that ongoing updates and active community support are crucial for long-term viability. These systemic challenges directly inform our strategic decision to focus on leveraging and extending established ontologies rather than immediately proposing a new, comprehensive DR ontology based on our identified IRs. Such an endeavor risks contributing to the very fragmentation and adoption difficulties that hinder many specialized ontologies. Consequently, our research centers on Brick, which is a widely accepted ontology with strong research community backing \cite{balaji2018brick}, industry adoption \cite{FierroPauwels2022}, and alignment with the upcoming ASHRAE Standard 223P \cite{open223intro}. We consider Brick in combination with the DR-focused ontologies like DELTA and EFOnt and a grid-specific model, CIM,  to address remaining specific DR-related entities and grid-related requirements. We believe that most IRs identified in this work can be effectively addressed by enhancing these existing frameworks, with concrete extensions proposed in Section \ref{sec:extensions}. 
While acknowledging that highly specific IRs would require development of new ontologies, our primary methodology aims to build upon existing community efforts and foster broader interoperability rather than contributing to further fragmentation in the landscape of ontology research.

Recent research also underscores the need for portable demand flexibility applications \cite{paul2023open, de2024enabling}. For example, a framework employing Brick has been used for price-based DR strategies during the control stage, relying on configuration templates to address representational gaps \cite{de2024enabling}. Expanding on this work, OpenBOS was developed to validate and semi-automatically configure applications in line with ASHRAE 223 standards \cite{paul2023open}. Our study distinguishes itself from these developments in several key areas: (1) We focus on incentive-based DR, which requires additional computational complexities due to restrictions from ISOs and higher upfront setup costs, which underscore the importance of interoperability; (2) While \cite{de2024enabling} focus on controls, we delve into the wholesale DR business process, covering everything from enrollment and qualification to scheduling, notification, deployment, real-time communications, and performance measurement; (3) Instead of creating a control interface, our focus is on improving the availability and flow of information necessary for each stage of DR participation. 

\begin{figure}[h]
\centering
\includegraphics[width=0.8\columnwidth]{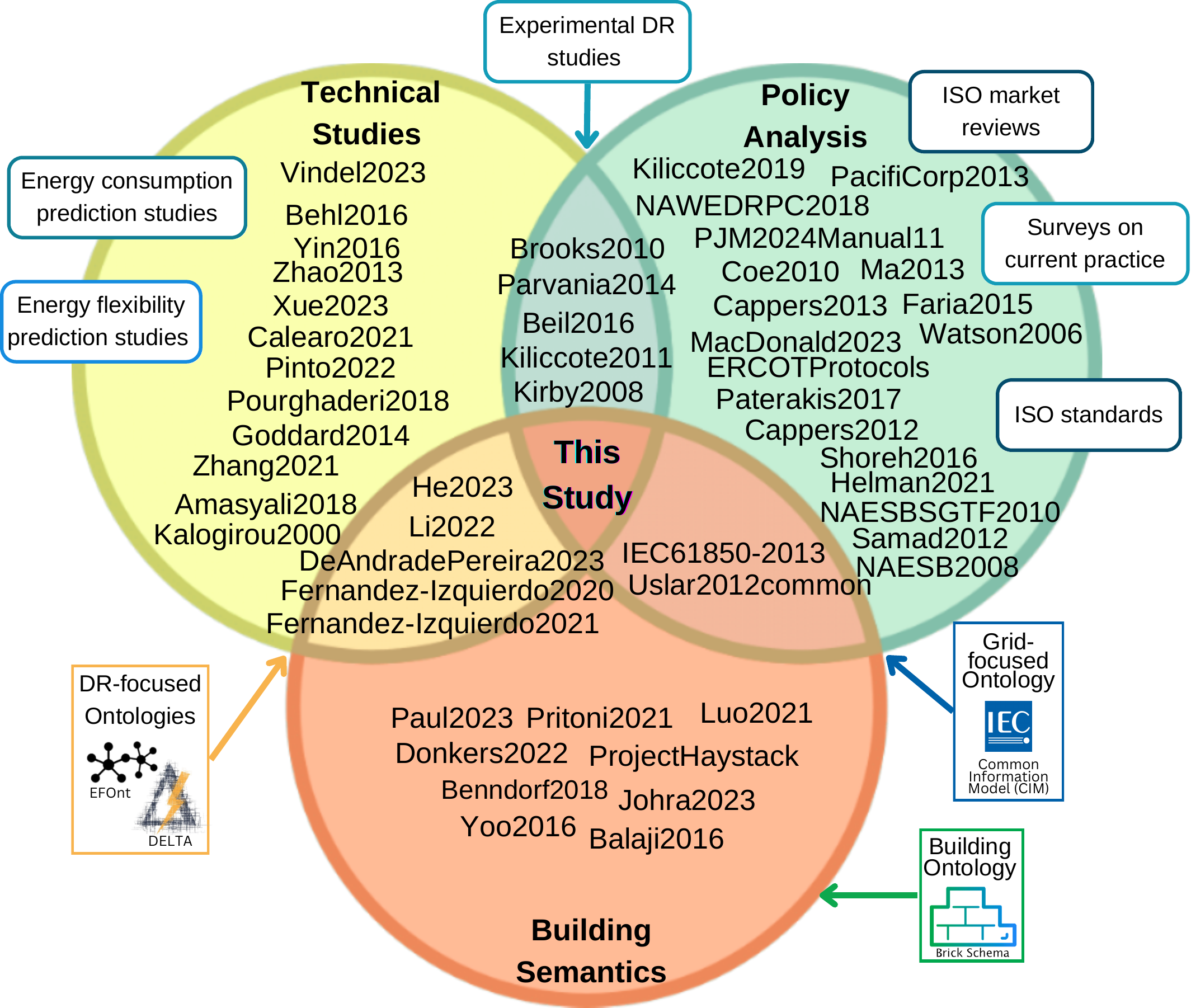}
\caption{Illustration of the intersection of Technical Studies, Policy Analysis, and Building Semantics in the context of energy flexibility research and DR policy.}
\label{fig:overview}
\end{figure}

This paper is devoted to outlining the comprehensive array of IRs pivotal for DR in commercial buildings, bridging the gap between technical research, existing policies, and semantic technologies. An overview of the existing landscape of research including the studies used for extracting the IRs is illustrated in Figure \ref{fig:overview}. To our knowledge, ours is the first attempt to rigorously define IRs for ontology development at the intersection of commercial buildings and DR, highlighting a novel perspective in semantic ontology research.

\vspace{-3ex}
\section{Methodology and Data Collection} \label{sec:ontology}
To identify the IRs necessary for semantic interoperability, we adopted a process-centric approach, analyzing the end-to-end business lifecycle of DR. We decomposed this lifecycle into four distinct stages to ensure comprehensive coverage of both commercial and technical necessities. For each stage, we conducted a comprehensive review of diverse data sources, including manuals from ISOs such as PJM \cite{PJM2024Manual11, PJM2024DemandResponse}, standards from the NAESB \cite{NAESB2008, NAESBSGTF2010}, and academic studies selected using the specific keywords presented in Figure \ref{fig:questions}. This multi-source review allowed us to extract the specific data elements required for decision-making, execution, and verification. The resulting IRs, a subset of which is shown in Table \ref{tab:ir_examples_stages} (full table in Appendix Table \ref{table:ir_class}), form the empirical basis for the ontology coverage analysis presented in subsequent sections.

\subsection{Identification of Information Requirements}

To systematically identify the data dependencies for automating DR, we analyzed the workflow across four distinct stages. Table \ref{tab:ir_examples_stages} provides example IRs identified for each stage. In the following sections, we detail the specific applications that drive these data needs and identify the primary stakeholders—grid operators (ISOs), aggregators (CSPs/utilities), and building operators—who utilize them.

The IRs for any given building or flexible asset are conditional on the specific DR program and the existing equipment the building possesses. Since there are multiple ways to participate in DR and reduce energy consumption, buildings will leverage the systems and data points they already have available. While simpler participation strategies may only require basic telemetry, we deliberately include more complex approaches identified in the literature (such as damper positions and electric vehicle charging status) to increase the comprehensiveness of our list.

\begin{figure}[h]
\centering
\includegraphics[width=1\columnwidth]{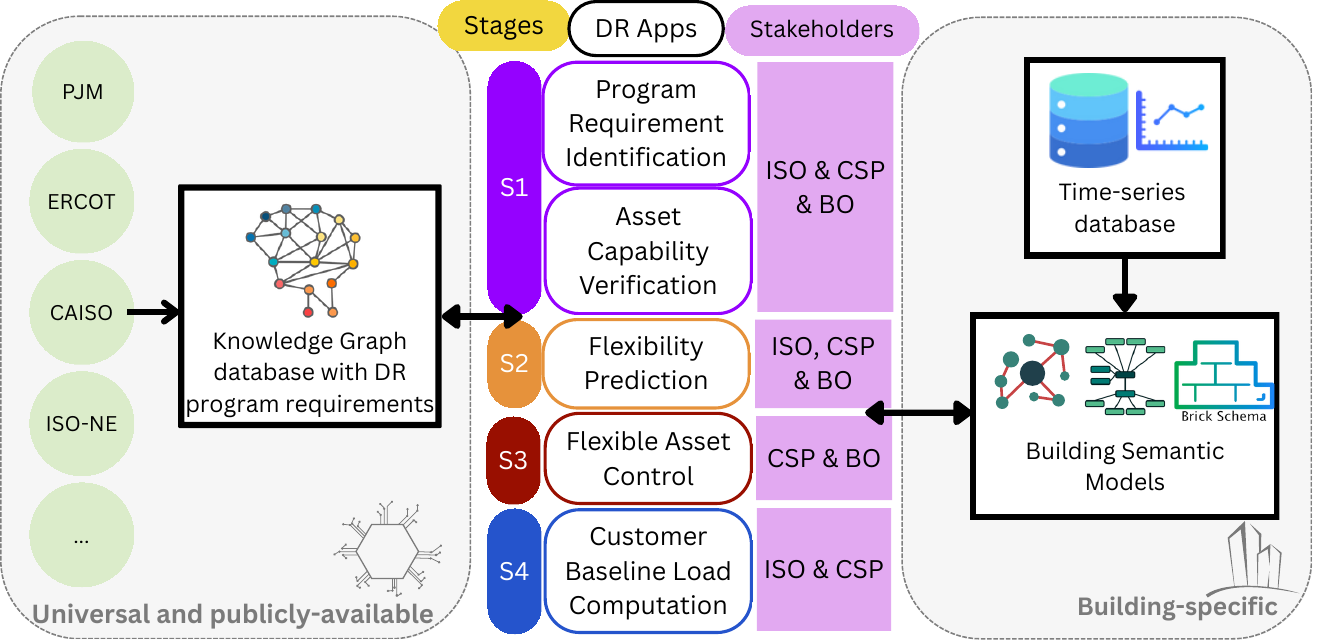}
\caption{Visualization of bridging market requirements and building semantic models via stage-specific applications.}
\label{fig:semantic-market}
\end{figure}

\begin{table}[h!]
\caption{Example Information Requirements  for Each Stage of DR Participation.}
\label{tab:ir_examples_stages}
\centering
\scalebox{0.95}{
\begin{tabular}{l}
\hline
\textbf{1. Enrollment \& Qualification} \\
- Minimum Eligible Resource Size \cite{NAESB2008, NAWEDRPC2018} \\
- Meter Interval \cite{NAESB2008, NAWEDRPC2018} \\
- Meter Accuracy \cite{NAESB2008, NAWEDRPC2018} \\
- Telemetry Reporting Interval \cite{NAESB2008, NAWEDRPC2018} \\
\hline
\textbf{2. Scheduling \& Award Notification} \\
- Outside air temperature \cite{yin2016quantifying, behl2016data, PJM2024Manual11, pinto2022towards} \\
- Zone air temperature setpoint \cite{vindel2023alphashed, behl2016data, kiliccote_characterization_2019} \\
- HVAC baseline load predictions \cite{vindel2023alphashed} \\
- Outside air temperature forecast \cite{yin2016quantifying, behl2016data} \\
\hline
\textbf{3. Deployment \& Real-time Communications} \\
- Fan power (VFD) \cite{beil_frequency_2016} \\
- Occupancy schedule \cite{vindel2023alphashed, behl2016data} \\
- DR start and end signal / Price level \cite{beil_frequency_2016, kiliccote_characterization_2019} \\
- EV Charging status \cite{zhao_integrated_2013} \\
\hline
\textbf{4. Measurement \& Performance Evaluation} \\
- X and Y values (for HighXofY, defined by ISOs) \cite{johra2023iea, PJM2024Manual11} \\
- Building electricity consumption \cite{vindel2023alphashed, yin2016quantifying, johra2023iea, xue2023utilizing, PJM2024Manual11, pinto2022towards} \\
- Outside air temperature \cite{yin2016quantifying, behl2016data, PJM2024Manual11, pinto2022towards} \\
- Day of the week / Time of the day \cite{behl2016data, johra2023iea, xue2023utilizing, PJM2024Manual11, pinto2022towards} \\
\hline
\end{tabular}
}
\vspace{-3ex}
\end{table}

\subsubsection{Enrollment \& Qualification} The process begins with enrollment and qualification. This stage is performed once to establish eligibility, while the subsequent stages are repeated throughout the participation. In this stage, CSPs and building operators would utilize two fundamental applications to navigate market entry. First, they perform \textbf{Program Requirement Identification} to query the specific regulatory and technical prerequisites established by the ISO. Second, they perform \textbf{Asset Capability Verification}, where the building operator inventories their physical systems and the CSP assesses if those assets satisfy requirements such as \textit{minimum eligible resource size} or specific \textit{telemetry reporting intervals}. We illustrate the workflow and data dependencies between these two applications in Figure \ref{fig:semantic-market}. This stage is dominated mostly by regulatory constraints. Our review utilized foundational standards from NAESB \cite{NAESBSGTF2010} and surveys on participation barriers \cite{shoreh_survey_2016} to motivate these applications. The identified IRs are a subset of the definitions established by NAESB \cite{NAESB2008}, which was also used in \cite{NAWEDRPC2018}. These definitions encompass critical technical parameters, such as \textit{meter accuracy} and \textit{meter interval}, which CSPs and/or building operators must rigorously validate against the ISO's standards to ensure eligibility. A detailed analysis is provided in Appendix \ref{sec:enrollment}. It is important to note that while there are additional IRs that are necessary to be modeled based on the communication protocol being used across the stakeholders (such as IEEE 2030.5 and OpenADR), that is outside of scope for this paper.

\subsubsection{Scheduling \& Award Notification} Once enrolled in a program, the focus shifts to scheduling and award notification. The primary application in this phase is \textbf{Flexibility Prediction}. This is a collaborative application: the building operator (or their CSP) estimates the load curtailment potential for Day-Ahead and Intra-Day horizons to avoid occupant discomfort, while the CSP aggregates these individual predictions to formulate a compliant bid for the ISO \cite{yin2016quantifying, liu2022developing}. Because miscalculating this capacity can lead to significant financial penalties, the accuracy of this application is critical for both the CSP (managing risk) and the building operator (managing operations). We examined prediction methodologies ranging from models based on \textit{outside air temperature} and \textit{outside air temperature forecasts} \cite{yin2016quantifying} to physics-informed approaches using VAV damper positions like AlphaShed \cite{vindel2023alphashed}. Our synthesis highlights a dichotomy: while industry often relies on simple weather correlations, existing studies demonstrate that high-fidelity predictions require granular building data, such as \textit{zone air temperature setpoints} or VAV damper positions \cite{vindel2023alphashed, behl2016data, kiliccote_characterization_2019}. Therefore, the IRs for this stage rely heavily on deep operational data and \textit{HVAC baseline load predictions}. The detailed extraction of IRs for this stage is available in Appendix \ref{sec:scheduling}.

\subsubsection{Deployment \& Real-Time Communications} The deployment stage requires a clear separation of concerns, though the locus of control varies by the type of service. The process is triggered by the ISO, which issues the dispatch instruction containing parameters like the \textit{DR start and end signal / price level}. The CSP typically receives this signal. The building must then translate this high-level signal into granular \textbf{Flexible Asset Control} logic, determining how to modulate specific systems. For fast-response automated processes like regulation, the CSP often bypasses the building's central logic to directly actuate specific systems \cite{beil_frequency_2016}. Conversely, for longer-duration services, the building operator may have operational authority, translating the high-level signal into granular control logic using strategies such as Global Temperature Adjustment (GTA) \cite{behl2016data}. 
In both cases control decisions for the flexible assets responding to the DR signals typically occur within supervisory control loops (minute-to-hour timescales). At this level, ontologies like Brick provide observability, enabling control applications to discover actuators, data sources, network addresses, and time-series references for flexible assets. Low-level control loops that are not specific to a DR-strategy fall outside this scope of the IR review.
To capture the full range of data needs for both direct-control and operator-mediated scenarios, we analyzed strategies for modulating components like fan power (VFD) and regulation services \cite{beil_frequency_2016}. Synthesizing these sources reveals that stage 3 IRs significantly overlap with stage 2 IRs, particularly regarding \textit{occupancy schedules}, but with added temporal urgency. Furthermore, the integration of EVs introduces unique variables such as \textit{EV charging status} \cite{zhao_integrated_2013}, representing a new category of mobile asset data that building operators must integrate into static management systems. The IRs identified for these diverse strategies are detailed in Appendix \ref{sec:deployment}.
As mentioned earlier, the IRs in this stage depends on the existing capabilities, available data and control points of a building and its systems.

\subsubsection{Measurement \& Performance Evaluation} The final stage is measurement and performance evaluation. The key application in this phase is \textbf{Customer Baseline Load Computation}. This application is primarily executed by the ISO to determine financial settlements and verify performance, though CSPs frequently run parallel calculations to audit these settlements. The application utilizes historical metering data, specifically \textit{building electricity consumption}, to calculate the CBL—an estimate of what consumption would have been absent the DR event. Our review contrasted industry-standard averaging techniques, which rely on defined \textit{X and Y values} (for HighXofY methods\footnote{It uses Y similar days to the event day and takes the highest X of them for averaging \cite{johra2023iea}}) \cite{PJM2024Manual11}, as well as research studies focusing on data-driven approaches. These included regression trees \cite{behl2016data}, extreme gradient boosting methods \cite{pinto2022towards}, and even novel applications of language models like Bart and Pegasus \cite{xue2023utilizing}. Unlike the standard averaging methods, these advanced models necessitate a broader set of inputs, specifically requiring correlated environmental data like \textit{outside air temperature} and temporal context such as \textit{day of the week / time of the day}. This expansion of input variables highlights a primary use case for semantic interoperability—standardizing these diverse definitions to ensure the models can be consistently applied across different building systems by ISOs, CSPs, and Building operators. The detailed review is found in Appendix \ref{sec:measurement}.

\subsection{Synthesis of Information Requirements}
To systematically analyze the capacity of ontologies to meet the needs of DR applications, we synthesized and categorized IRs into conceptual categories. These categories provide a structured representation of the diverse requirements across DR stages, offering a high-level view of the ontologies' capabilities. Please refer to the first column of Table \ref{tab:ontology_roadmap} for the list of these IR categories; further details and a comprehensive classification can be found in Table \ref{table:ir_class} in the Appendix. Some IRs appear frequently across studies—for example, past DR schedules and building electricity consumption—which are essential for training flexibility forecasting models and predicting baseline loads. Time-based parameters are another recurring theme, owing to the dynamic nature of HVAC systems, often requiring temporal considerations to address fluctuations. Additionally, outdoor air temperature significantly impacts HVAC energy consumption, while zone air temperature setpoints play a critical role in flexibility prediction algorithms and GTA analyses. 

\begin{figure}[h]
\centering
\includegraphics[width=1\columnwidth]{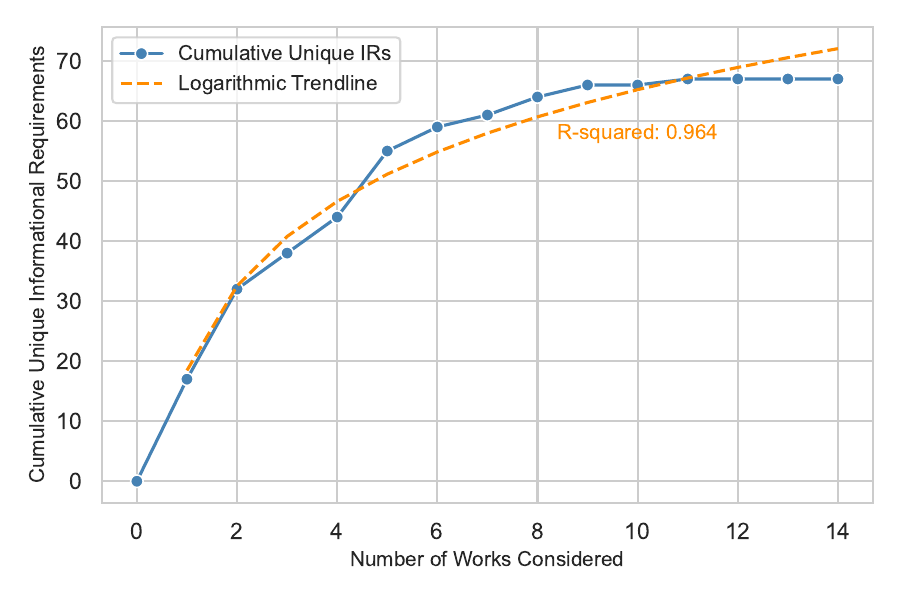}
\caption{Cumulative unique information requirements vs. number of works considered, with a logarithmic trendline showing diminishing returns as additional works contribute fewer unique IRs.}
\label{fig:returns}
\end{figure}

The validity of these IRs is demonstrated in two ways. First, Table \ref{table:ir_class} summarizes the frequency of specific IRs in the sources analyzed, highlighting their relevance and prevalence. Second, we visualized the diminishing returns in the discovery of unique IRs as additional studies were reviewed, shown in Figure \ref{fig:returns}. Figure \ref{fig:returns} plots the cumulative number of unique IRs identified as the number of reviewed studies increases. The blue line represents the actual cumulative count of unique IRs, while the orange dashed line represents a logarithmic trendline fitted to the data. The logarithmic trendline demonstrates the diminishing rate at which new IRs are discovered with each additional study. Its high $R^2$ value (0.964) indicates a strong fit, reflecting that the review process is reaching a point of saturation. Beyond this point, additional studies contribute minimal new IRs, emphasizing the comprehensiveness of the review. This pattern confirms that the identified IRs effectively capture the majority of relevant requirements. Together, the quantitative analysis in Table \ref{table:ir_class} and the diminishing returns illustrated in Figure \ref{fig:returns} validate the robustness and comprehensiveness of the identified IRs. 

\subsection{Overview of Selected Ontologies} Before proceeding with the coverage analysis, we briefly introduce the ontologies under consideration to contextualize their scope and limitations. Brick \cite{balaji2016brick} is an ontology widely adopted for representing commercial building data. We prioritized Brick over alternatives like Project Haystack \cite{ProjectHaystack}, W3C SSN/SOSA (Semantic Sensor Network / Sensor, Observation, Sample, and Actuator)~\cite{SSN_SOSA_2017} and SAREF \cite{etsi_saref_2014} for specific strategic reasons. Unlike Haystack 3.0, which primarily relies on a semi-structured tagging system, Brick employs a formal, class-based ontology structure. This formalization provides the strict semantic constraints and relationship definitions necessary to model the complex interdependencies of DR IRs—a level of rigor that is harder to enforce with a purely tag-based approach. Furthermore, Brick benefits from strong research community backing \cite{balaji2018brick}, demonstrated industry adoption \cite{FierroPauwels2022}, and strategic alignment with the ASHRAE 223 initiative for standardizing semantic ontologies \cite{open223intro}. 
SSN/SOSA is a foundational observation meta-ontology that describes how measurements are made and by what sensor, rather than defining domain-specific energy system concepts. Every IR in our catalog would require custom domain extensions built on top of SSN/SOSA, yielding no direct coverage of DR-specific entities.
Similarly, while SAREF is prominent in Europe, its limited adoption in the US market renders it less practical for evaluating the DR programs in the US that form the basis of this study. Consequently, while Brick serves as a robust foundation for physical assets, its scope does not explicitly cater to DR applications, limiting its ability to represent domain-specific entities such as regulatory constraints or DR management.

The OpenADR Ontology extends the OpenADR2 protocol standards through semantic enrichment, integrating publicly available ontologies such as OWL-time for temporal conditions and GeoSPARQL for geospatial data \cite{fernandez2020openadr}. DELTA, an extension of the OpenADR Ontology \cite{fernandez2021supporting}, builds upon this foundation by incorporating SAREF for expressing properties and measurements and SAREF4CITY for modeling KPIs. Given their close relationship, we will consider the OpenADR Ontology and DELTA together under the name DELTA. While DELTA introduces capabilities for modeling smart homes and HVAC systems, its lack of formal definitions and categorizations limits it to high-level representations, making it insufficient for detailed DR actions in complex environments such as commercial buildings.

EFOnt is a recently developed ontology focused on energy flexibility, designed as part of the IEA Annex 81 initiative. Its primary purpose is to support KPI computation for energy flexibility applications \cite{li_semantic_2022}. Although EFOnt offers a specialized focus on energy flexibility, its coverage remains limited, often relying on specific frameworks like EnergyPlus for detailed modeling.

We also evaluated the CIM, a comprehensive abstract standard developed by the electric power industry and officially maintained by the International Electrotechnical Commission (IEC). For the scope of this study, we focused on three specific CIM standards highly relevant to DR operations: IEC 61970-301, which establishes the base physical topology and objects for energy management systems; IEC 61968-9, which defines interfaces for distribution management, specifically focusing on advanced meter reading and control; and IEC 62325-301, which provides the framework and extensions for deregulated energy market communications. Unlike Brick, DELTA, and EFOnt, the official CIM standards is not free. This closed nature presents a significant barrier to academic research, smaller aggregators, and community-driven application development. Consequently, our coverage evaluation of CIM was conducted using publicly available open-source representations, user-group schemas, and community documentation derived from the standard, rather than the IEC documents themselves~\cite{Zepben_CIM100, uslar2012common, PNNL34946} Despite its lack of open-source accessibility, CIM's widespread adoption by utilities and ISOs makes it an essential resource for evaluating the grid-facing and market-oriented requirements of DR programs. 
We had also shortlisted IEC 61850~\cite{IEC61850_2013}, but excluded it in our final evaluation. While an important grid communication standard that addresses real-time control and protection of substation equipment at the distribution infrastructure level does not extend to the customer-premises DR participation layers that are the focus of this study.

\subsection{Methodology for Ontology Coverage Evaluation}
Here, we describe the methodology used for verifying ontology coverage for a certain IR. In the context of evaluating IRs, certain IRs can be described using a combination of classes and/or relationships. To formalize this, let each IR ($\text{IR}_i$) be represented as a set of components ($C_i$), where these components may include classes, relationships or properties.

Let $B$, $D$, $E$ and $M$ represent the sets of components supported by the Brick, DELTA, EFOnt, and CIM ontologies, respectively. For a given ontology $O \in \{B, D, E, M\}$, the set of supported components is denoted as $S_O$. An ontology $O$ satisfies an IR ($\text{IR}_i$) if and only if all components of $\text{IR}_i$ are contained in the ontology's support set in some form (i.e., paraphrases are acceptable). Formally, the satisfaction function ($R$) can be defined as:
$$R(\text{IR}_i, S_O) =
\begin{cases}
1 & \text{if } C_i \subseteq S_O, \\
0 & \text{otherwise}.
\end{cases}$$

The determination of whether $C_i \subseteq S_O$ (i.e., an IR is `satisfied') was performed manually by the authors, who rigorously reviewed the available classes and relationships in each ontology. This manual approach allowed for the practical interpretation of IR components being present `in some form' (as per the preceding definition where paraphrases are noted as acceptable), an interpretation that necessarily involved understanding the unique descriptive and semantic conventions of each individual ontology, especially for concepts within overlapping domains. Specifically, the authors considered: 1)~direct matches, where an IR component term was identical to an ontology component; 2)~synonyms, where commonly accepted synonyms were deemed equivalent (e.g., an IR component `Power Meter' could be matched by an ontology concept like \texttt{Electricity\_Meter}); and 3)~shapes, where an IR's full set of components ($C_i$) was satisfied by a specific combination of classes and/or relationships in the ontology (e.g., the IR `EV energy consumption' is satisfied if its constituent concepts, such as \texttt{ElectricVehicle} and \texttt{GenericLoadProfile} as later exemplified, are present and appropriately relatable). While a formal workload assessment for each IR evaluation was not conducted, the overall process of evaluating approximately 70 IRs across the four ontologies was considered manageable by the authors. Although an automated approach was beyond the scope of this study, the process has inherent value and replicability, suggesting that for a larger number of IRs, automated workflows leveraging natural language processing techniques could be developed.

For each ontology's components $S_O$ and each IR category $\mathcal{C}$, the percentage of coverage is computed as the fraction of satisfied IRs within that category, converted to a percentage. Let $\mathcal{I}_\mathcal{C}$ denote the set of IRs in category $\mathcal{C}$. The percentage of coverage for ontology $S_O$ over IR category $\mathcal{C}$, denoted as $\kappa_{S_O, \mathcal{C}}$, is computed as:
$$\kappa_{S_O, \mathcal{C}} = \left( \frac{\sum_{\text{IR}_i \in \mathcal{I}_\mathcal{C}} R(\text{IR}_i, S_O)}{|\mathcal{I}_\mathcal{C}|} \right) \times 100,$$
where:
\begin{itemize}
    \item $R(\text{IR}_i, S_O) = 1$ if $\text{IR}_i$ is satisfied by ontology components $S_O$, and $0$ otherwise,
    \item $|\mathcal{I}_\mathcal{C}|$ is the total number of IRs in category $\mathcal{C}$.
\end{itemize}

For the combination of the four ontologies $S = \{S_B, S_D, S_E, S_M\}$, the percentage of \textit{combined coverage} over an IR category $\mathcal{C}$, denoted as $\kappa_{S, \mathcal{C}}$, is computed as:

\begin{equation}
    \kappa_{S, \mathcal{C}} = \frac{\sum_{\text{IR}_i \in \mathcal{I}_\mathcal{C}} \max\left(R(\text{IR}_i, S_B), R(\text{IR}_i, S_E), R(\text{IR}_i, S_D), R(\text{IR}_i, S_M)\right)}{|\mathcal{I}_\mathcal{C}|} \times 100
\end{equation}

where $\max\left(R(\text{IR}_i, S_B), R(\text{IR}_i, S_E), R(\text{IR}_i, S_D), R(\text{IR}_i, S_M)\right)$ returns $1$ if at least one of the ontologies satisfies $\text{IR}_i$, and $0$ otherwise. Crucially, this method of determining combined coverage, by relying on the independent evaluation of each ontology's ability to satisfy an IR ($R(\text{IR}_i, S_O)$ as detailed previously), does not require a formal merging or alignment of the individual ontologies. Thus, potential logical conflicts or semantic inconsistencies that might arise if one were to attempt such a merge do not impede the calculation or interpretation of $\kappa_{S, \mathcal{C}}$. The goal is to demonstrate the combined coverage from the independent abilities of the ontologies, as the task of actually merging them was considered outside the scope of this work.

To further explain, we present two examples. For the first example, the IR \textit{EV energy consumption} is represented by the concepts \texttt{ElectricVehicle} and \texttt{GenericLoadProfile}, which are both present in ontology $S_E$. Since these concepts are missing in $S_B$, $S_D$, and $S_M$, only $S_E$ satisfies the IR, resulting in $R(\text{IR}, S_E) = 1$, while $R(\text{IR}, S_B) = 0$, $R(\text{IR}, S_D) = 0$, and $R(\text{IR}, S_M) = 0$. Using the combined coverage formula, $\max(0, 1, 0, 0) = 1$, the IR is satisfied in the combined coverage.

For the second example, the IR \textit{battery capacity} is represented by the concepts \texttt{Battery} and \texttt{StorageCapacity}, both found in the ontology $S_D$. These concepts are not present in $S_B$, $S_E$\, or $S_M$, so $R(\text{IR}, S_D) = 1$, while $R(\text{IR}, S_B) = 0$, $R(\text{IR}, S_E) = 0$, and $R(\text{IR}, S_M) = 0$. The combined coverage formula yields $\max(0, 0, 1, 0) = 1$, indicating that the IR is satisfied in the combined coverage.

\section{Results and Extensions}

This section evaluates the coverage of the defined IRs by current ontologies. Based on these findings, we propose a roadmap of targeted extensions and integrations to establish a unified semantic framework for DR.

\subsection{Coverage Assessment of Current Ontologies}
\begin{figure}[h]
\centering
\includegraphics[width=\columnwidth]{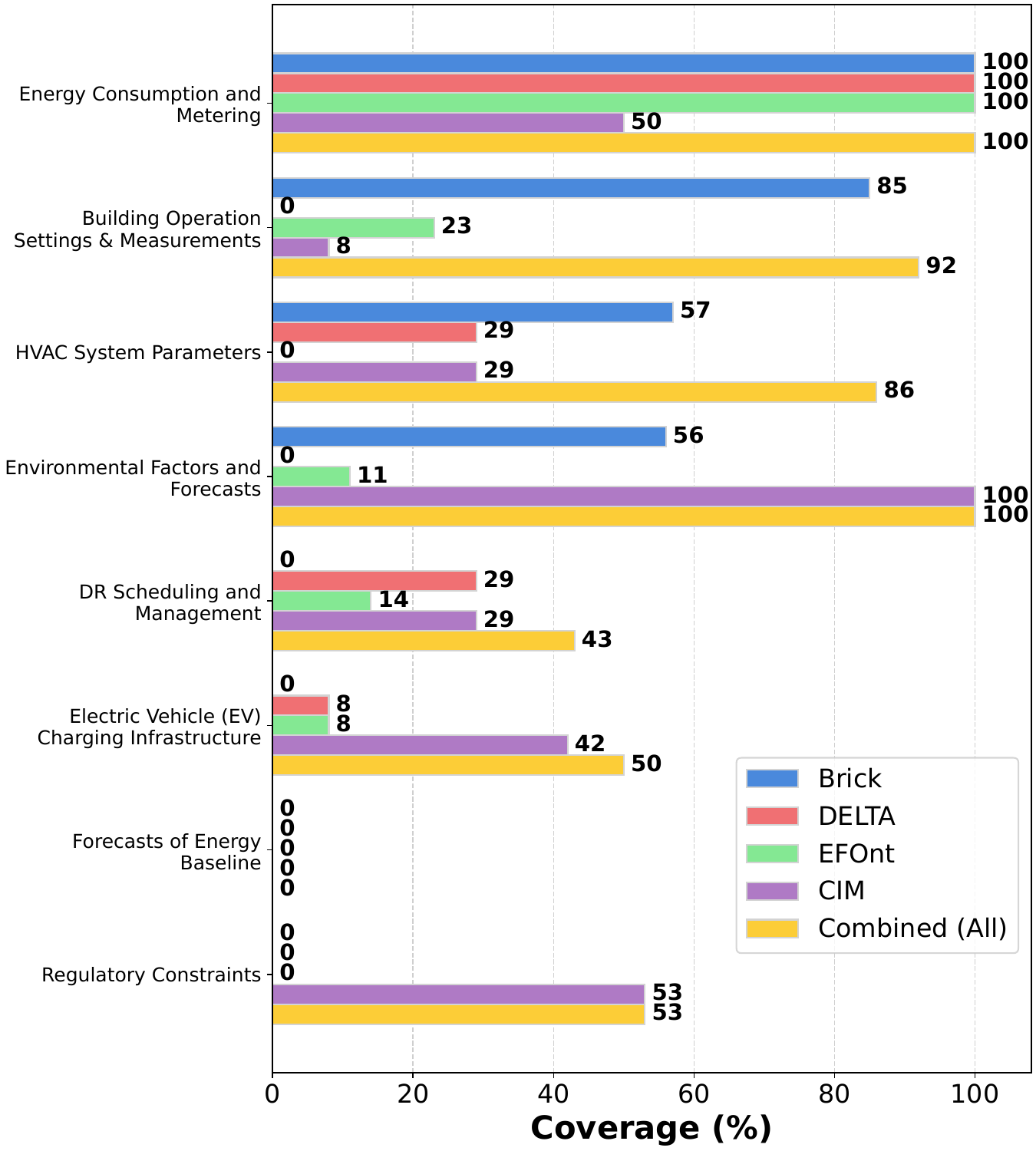}
\caption{Visualization of percent coverage of each IR category. Combined is computed as explained with $\kappa_{S_O, \mathcal{C}}$.}
\label{fig:vis_ont_coverage}
\vspace{-3ex}
\end{figure}

Figure \ref{fig:vis_ont_coverage} provides a detailed analysis of how well Brick ($B$) and DR-focused ontologies \( \{D,E\}\) meet various IRs\footnote{Coverage of DELTA also includes OpenADR Ontology since it was built on it.}. The x-axis demonstrates the \(\kappa_{S_O, \mathcal{C}}\) values for each ontology and for each IR category (shown on the y-axis) and the combined coverage across all ontologies \(\kappa_{S, \mathcal{C}}\).

This evaluation reveals significant gaps in the ability of semantic technologies, such as Brick and DR-focused ontologies, to meet the information needs of energy flexibility in buildings. Key deficiencies include inadequate support for regulatory constraints, weather forecasting, and the integration of emerging technologies like EV charging (see Appendix \ref{sec:roadmap} for further details). For instance, while \textbf{Brick} is effective for modeling building operations, it lacks support for dynamic weather factors and regulatory constraints, which are critical for adapting DR strategies to real-world conditions. For example, it misses IRs like \textit{Outside air temperature forecast} and regulatory parameters such as \textit{Minimum Eligible Resource Size} or \textit{Meter Interval}. It is important to note that these regulatory constraints were mostly derived from the ISO/RTO Council's 2018 effort to address market variety by cataloging DR program features across North America \cite{NAWEDRPC2018}. \textbf{DELTA and EFOnt} offer similarly partial coverage; although DELTA provides additional support for HVAC parameters and DR scheduling, both ontologies address only 29\% of DR scheduling and management IRs and fail to represent regulatory requirements entirely.

Beyond individual ontology gaps, the analysis identifies several critical deficiencies and findings common to the evaluated frameworks:
\begin{itemize}
\item None can represent the concept of forecasts for energy baselines, although CIM can do so for for weather conditions.
\item Emerging technologies, such as EV charging, are only partially represented, with combined coverage reaching 50\% largely due to CIM's electrical modeling, up from 16\% without it.
\item While Brick, DELTA, and EFOnt fail to support regulatory constraints (defined by the ISO/RTO Council's 2018 effort to address market variety by cataloging DR program features across North America \cite{NAWEDRPC2018}), CIM provides essential but incomplete coverage (53\%) for these market rules.
\item DR scheduling and management concepts are primarily tied to specific frameworks (e.g., EnergyPlus for EFOnt, OpenADR for DELTA, IEC 61968-9 for CIM), limiting their applicability to broader incentive-based DR methods.
\item In contrast to these gaps, energy consumption and metering are represented with 100\% coverage.
\end{itemize}

The analysis of IR coverage across DR stages reveals increasing complexity and data integration needs as the processes progress. Early stages like enrollment and qualification rely heavily on foundational IRs, such as regulatory constraints and initial system settings. In contrast, later stages, including deployment, real-time communications, and performance measurement, demand more detailed and dynamic data, such as environmental forecasts, energy metering, and building operation parameters. The limited coverage of these critical IRs by existing open-source ontologies underscores the need for a more robust and flexible framework, potentially bridging CIM's grid-level architecture with building-level semantics.

\begin{table*}[ht]
\centering
\caption{Roadmap of Ontology Extensions and Integrations for DR Automation.}
\vspace{-4mm}
\label{tab:ontology_roadmap}
\resizebox{\textwidth}{!}{%
\begin{tabular}{@{}l l p{5.5cm} p{8cm}@{}}
\toprule
\textbf{IR Category} & \textbf{Gap / Requirement} & \textbf{Proposed Class / Property} & \textbf{Source / Integration Strategy} \\ \midrule

\textbf{Regulatory Constraints} 
& ISO-market requirements & \texttt{cim:ResourceCapacity}, \texttt{cim:LoadBid} & \textbf{CIM / New Ontology}: Extend CIM or build an open-source ontology. \\ \midrule

\multirow{3}{*}{\textbf{HVAC System Parameters}} 
 & Telemetry & \texttt{brick:Telemetry}, \texttt{oadr:TelemetryReport} & \textbf{Extend Brick/Integrate OpenADR}: Create a Brick telemetry class or use OpenADR instance. \\
& Telemetry/Meter Interval & \texttt{oadr:hasSamplingRate}, \texttt{cim:MeasuringPeriodKind} & \textbf{Extend Brick}: Apply OpenADR or CIM properties to meter/telemetry instances. \\
& Telemetry/Meter Accuracy & \texttt{oadr:hasAccuracy}, \texttt{cim:sensorAccuracy} & \textbf{Extend Brick}: Apply OpenADR  or CIM properties to meter/telemetry instances. \\ \midrule

\multirow{2}{*}{\textbf{\begin{tabular}[c]{@{}l@{}}Building Operation Settings \\ \& Measurements\end{tabular}}} 
 & Future Schedules & \texttt{Occupancy\_Schedule\_of\_Zones} & \textbf{Extend Brick}: Create class for future schedules. \\
& Price Automation & \texttt{Price\_Setpoint\_Matching}, \texttt{cim:LoadReductionPriceCurve} & \textbf{Extend Brick}: Link prices to setpoint adjustments using CIM reference. \\ \midrule

\multirow{7}{*}{\textbf{\begin{tabular}[c]{@{}l@{}}Demand Response Scheduling \\ \& Management\end{tabular}}} 
 & Event Modeling & \texttt{oadr:Event} & \textbf{Integration (OpenADR)}: Use existing event structure. \\
 & Event Timing & \texttt{hasStartingTime}, \texttt{hasEndingTime} & \textbf{Extension}: Fix ambiguity in \texttt{oadr:schedule} linking. \\
 & Performance Promise & \texttt{Reduction} & \textbf{Extension}: New entity for energy curtailed (missing in DELTA). \\
 & Historical Data & \texttt{Past\_DR\_Schedule} & \textbf{Extension}: Store historical event logs. \\
 & Pre-conditioning & \texttt{efont:pre\_heating}, \texttt{efont:pre\_cooling} & \textbf{Integration (EFOnt)}: Adopt thermal system concepts. \\
 & Control Logic & \texttt{Setpoint\_Change} & \textbf{Extension}: Link EFOnt concepts to specific setpoint actions. \\ \midrule

\multirow{3}{*}{\textbf{EV Charging Infrastructure}} 
 & EV Asset & \texttt{Electric\_Vehicle} & \textbf{Extend Brick}: Add native EV class (EFOnt's class is designed for simulation). \\
& EV Storage & \texttt{delta:Battery}, \texttt{delta:StorageCapacity}, \texttt{cim:VoltageLimit} & \textbf{Extend Brick}: Integrate DELTA and CIM models for battery limits. \\
 & Load Profile & \texttt{efont:GenericLoadProfile} & \textbf{Extend Brick}: Integrate EFOnt model for consumption profile. \\ \midrule

\textbf{\begin{tabular}[c]{@{}l@{}}Energy Consumption \\ \& Metering\end{tabular}} 
 & Submetering (Thermal) & \texttt{Coefficient\_Of\_Performance} & \textbf{Extend Brick}: Add conversion factor for thermal-to-electric calculation. \\ \midrule

\textbf{Time Based Parameters} 
 & Holidays & \texttt{Off\_Business\_Schedule} & \textbf{Extend Brick}: Model non-standard days for baselining. \\ \midrule

\multirow{3}{*}{\textbf{\begin{tabular}[c]{@{}l@{}}Forecasts of Energy Baseline \\ \& Environmental Factors\end{tabular}}} 
 & Baseline Output & \texttt{Building\_Baseline\_Prediction} & \textbf{Extend Brick}: Store model outputs distinct from meters. \\
 & HVAC Output & \texttt{HVAC\_Baseline\_Prediction} & \textbf{Extend Brick}: Store model outputs for sub-systems. \\
& Forecast Context & \texttt{hasForecastReference}, \texttt{cim:Forecast} & \textbf{Extend Brick}: Handle multiple forecast generation times mimicking CIM's architecture. \\ \bottomrule
\end{tabular}%
}
\vspace{-2ex}
\end{table*}
\subsection{Proposed Ontology Extensions and Integrations} 
\label{sec:extensions}

To address these gaps, we propose a framework for enhancing Brick through strategic extensions and integrations with DR-focused ontologies and utility standards. The decision to extend Brick is guided by the principle of aligning IRs within its scope, such as HVAC system parameters and building operation settings, with its existing infrastructure. IRs outside Brick's scope are better addressed through integrations with complementary ontologies like an ISO/RTO-market Ontology, CIM, and DR-related Ontologies such as DELTA and EFOnt.

In our envisioned framework, Brick serves as the central ontology, extended to cover its scoped IRs and integrated with external ontologies for specialized applications. Similarly, prior work has demonstrated potential of linking Brick with an energy storage ontology \cite{he2023ontology}. Additionally, collaborations like the IEA Annex-81 Community's work on Data-Driven Smart Buildings and Building-to-Grid Applications highlight the broader potential for high-level integrations, including those between Brick and EFOnt \cite{johra2023iea}. Table \ref{tab:ontology_roadmap} summarizes the specific class and property modifications required across the domains. In the following subsections, we describe the extensions needed at a high-level; for the detailed rationale and integration mechanisms, we refer the readers to the Appendix \ref{sec:roadmap}.

\subsubsection{Regulatory Constraints and External Integrations} Neither Brick nor DR-focused ontologies currently have the capability to represent regulatory constraints effectively as they fall out of scope. While CIM successfully covers approximately half of these regulatory constraints, a functional knowledge graph utilizing it for this purpose is not yet available. Therefore, we propose a future direction where these constraints are represented by a knowledge graph developed using a centralized ISO ontology, which requires either extending CIM or developing a new open-source alternative. Hosting this knowledge graph on an online platform (such as OpenEI~\cite{openei}), maintained by a committee of representatives from ISOs and major CSPs, would provide a unified source of information on program requirements, significantly lowering the barriers posed by market diversity. This platform should enable users to identify eligible programs and calculate expected revenue. While commercial tools like Enel North America's calculator \cite{enel_demand_response_calculator} exist for this purpose, they are limited to rough financial estimates based on simplified inputs. By leveraging the granular IRs defined in this study, a semantic-driven platform could go beyond simple approximations to validate technical eligibility against detailed site characteristics. Additionally, because certain IRs like minimum eligible resource size are difficult to calculate, future studies must focus on developing predictive methods to offer preliminary estimations of resource size for this computational step.

\subsubsection{Physical Asset Extensions: HVAC, EVs, and Metering} For physical assets, we analyzed how identified IRs could be modeled consistent with Brick’s existing structure. HVAC System Parameters are often fixed values; while Brick includes the property \texttt{value} for such assignments, it lacks relations for interval or accuracy metadata. We propose adopting classes from the OpenADR Ontology and CIM, such as \texttt{hasAccuracy}, \texttt{hasSamplingRate}, \texttt{sensorAccuracy}, and \texttt{MeasuringPeriodKind}, to model telemetry accuracy and reporting intervals within Brick. Regarding energy consumption and metering, Brick is able to represent the required IRs.

Furthermore, Brick currently lacks classes for EVs. While EFOnt includes an \texttt{ElectricVehicle} class, it focuses on modeling relations, and other specific EV ontologies focus on life cycle management \cite{yoo2016semantic}. Given that EV charging infrastructure is becoming a standard part of commercial buildings, the simplest solution is to extend Brick itself to include an \texttt{ElectricVehicle} class. This extension should support IRs related to performance, such as \texttt{StateOfCharge}, \texttt{ChargingVoltageLimit}, and \texttt{DischargeAllowance}. Brick can adopt CIM's established electrical entities to represent many of these parameters, which are currently missing from the other available schemas.

\subsubsection{Operational Extensions: Scheduling, Forecasts, and Management} Operational modeling requires bridging the gap between static assets and dynamic DR events. Previous studies identified that most DR classes were not available in Brick, necessitating self-configuration. We propose extending the concept of schedules to include specific DR event properties. While the OpenADR Ontology includes a \texttt{schedule} class, we recommend extensions such as \texttt{hasStartingTime} and \texttt{hasEndingTime}, or adopting CIM's \texttt{scheduledInterval}, to create a unified data storage method for accessing former DR event data. Additionally, while DELTA includes an entity for capacity, defined as the designated resource amount assigned to the grid that is compensated even if uncalled, it lacks an entity for Reduction. Reduction refers to the decrease in load promised within a period which is subject to penalties if not met, and thus these two distinct market products should be modeled separately.

For forecasting, we must distinguish model outputs from real-time measurements and account for overlapping predictions where multiple forecasts (generated at different times) target the same interval. We recommend introducing a property, \texttt{hasForecastReference}, applicable to classes like \texttt{Building\_Baseline\_Predictions} or environmental forecasts. This allows the system to uniquely index and retrieve these prediction sets in the database while maintaining semantic clarity. Finally, regarding pre-conditioning, we observed that while EFOnt links the definition of pre-heating strategies to physics-based simulations (e.g., EnergyPlus), real-life applications often rely on simple heuristic rules. We propose leveraging the \texttt{pre\_heating} or \texttt{pre\_cooling} concepts from EFOnt but enhanced with \texttt{hasStartingTime} and \texttt{hasEndingTime}, and a \texttt{Setpoint\_Change} relation. This formalizes the timing and scope of preconditioning actions without requiring a detailed EnergyPlus model for every building.

In summary, the results highlight the critical gaps in current ontologies in addressing essential IRs for achieving effective semantic interoperability in DR processes. The proposed extensions are designed not only to bridge these gaps but also to establish a robust foundation for future developments, ensuring adaptability to emerging technologies and evolving industry requirements.

\section{Conclusions}
\label{sec:conclusions}

In this paper, we have thoroughly reviewed the key IRs necessary to enable portable applications for commercial building DR participation, identifying necessary extensions and integrations for the existing semantic web ontologies. By mapping these requirements against the current landscape, we demonstrated that existing ontologies fall far short in representing the operational complexity required for deployment. While the combination of the leading commercial building ontology (Brick), specialized DR ontologies (DELTA, EFOnt) and a grid-specific ontology (CIM) provides a foundation for physical asset modeling, it remains fundamentally disconnected from the economic and regulatory realities of the grid. Specifically, our evaluation revealed deep systemic gaps: the inability to represent heterogeneous ISO program rules, and the lack of support for dynamic flexibility and weather forecasts.

Future research should focus on implementing the proposed extensions and evaluating their performance in diverse real-world buildings. Demonstrating the broad practical application and significance of our findings through a comprehensive case study, while beyond this paper's scope, is a key priority. However, to demonstrate feasibility, we have conducted an initial proof-of-concept implementation. As detailed in Appendix \ref{sec:alphashed_proof} and illustrated in Figure \ref{fig:alphashed23}, this preliminary effort applies the proposed extensions to a single real-world building for a single application (\cite{vindel2023alphashed}) using time-series data and historical DR events. While humble in scope, this single-building application serves as an initial first step toward demonstrating real-world deployment potential.

We acknowledge that, like the DR programs and building technologies they represent, ontologies require continuous revision to remain relevant. While our IR catalog reflects a comprehensive view of current and near-future practice, future work should revisit and extend these requirements as emerging technologies such as vehicle-to-grid and grid-edge storage mature, and as evolving ISO market structures introduce new information exchange needs. The framework presented in this paper supports both extending the requirements and also analyzing additional ontologies to model them as needed. 
However, the foundational methods and regulatory parameters identified in this work have remained structurally stable over time, in part because the semantic interoperability barriers this paper addresses have themselves slowed their broader adoption. Resolving these barriers is therefore both timely and necessary.

Ultimately, this paper aims to serve as a comprehensive guide for the continuous evolution of semantic technologies, with the goal of achieving portable and scalable DR applications, making it easier for building operators to participate in DR programs, and facilitating the translation of research findings into practical solutions.


\begin{acks}
Ozan Baris was supported by Pennsylvania Infrastructure Technology Alliance. This work was also supported by the Assistant Secretary for Critical Minerals and Energy Innovation, Office of Building Technology of the U.S. Department of Energy under Contract No. DE-AC02-05CH11231. Mitali Shah acknowledges that portions of the work presented in this paper were conducted during previous employment at JCI. Mitali is grateful for the technical environment and support provided during that period.
\end{acks}


\bibliographystyle{ACM-Reference-Format}
\bibliography{references}

\appendix

\section{Detailed Characterization of DR Services}\label{sec:appendix_services}

\textbf{Energy Services} are typically designed to address energy supply-demand imbalances. Participants often offer to adjust their load by bidding into day-ahead or intra-day markets (Stage 2). Transactions are generally based on the volume of energy reduction bid and subsequently delivered, with performance criteria—such as PJM's \cite{PJM2024Manual11} requirement for reductions to be within $\pm$20\% of the bid value—validated during the measurement phase (Stage 4 explained in Appendix \ref{sec:measurement}). For example, in commercial buildings, facility personnel might respond to a deployment dispatch signal (Stage 3) by increasing zone deadbands in unoccupied zones (See Appendix \ref{sec:deployment} for more details). 

\textbf{Capacity Services} are designed to ensure long-term grid reliability. These programs typically compensate participants based on a pre-agreed commitment to be available to reduce load (or provide capacity) over a longer term (e.g., a season or a year), often regardless of whether they are actually dispatched \cite{PJM2024DemandResponse}. For instance, a large data center may commit to be available to curtail 5 MW during peak summer months for a fixed payment. Given their infrequent use, which can be as rare as not occurring within a year \cite{PJM2024DemandResponse}, automation and application portability are likely to offer limited benefit in the context of capacity services.

\textbf{Ancillary Services} are crucial for maintaining immediate grid stability. Among these, Regulation services involve rapid adjustments to load within seconds to minutes. Participation typically necessitates rigorous telemetry installations verified during enrollment (Stage 1) \cite{paterakis_overview_2017}. Examples include large refrigeration plants momentarily turning off equipment via automatic switches (Stage 3), or resources like batteries or water heaters providing the necessary fast response \cite{beil_frequency_2016} (See Appendix \ref{sec:deployment} for more details). Reserve services ensure that adequate capacity is available to the system operator to manage unexpected events. These often require ramping up\footnote{We distinguish this physical ramping from the ``ramping reserves'' service defined in \cite{denholm2019introduction}. Here, ramping rate refers to how quickly a resource can start providing the promised load reduction.} within specified times (e.g., under 10 minutes for some reserves). Reserves are further segmented. Operating Reserves include spinning reserves (resources already synchronized to the grid and capable of responding automatically within minutes; for instance, a hotel demonstrated participation with responses between 12 to 60 seconds \cite{kirby2008spinning}) and non-spinning reserves (resources that can be brought online or curtailed, often via manual or semi-automated initiation, also within minutes, though DR participation in non-spinning reserves is not permitted by some ISOs like PJM \cite{PJM2024Manual11}). Secondary reserves (or supplemental reserves) provide a more sustained response, typically activating within 10 to 30 minutes and can be semi-automated or manual. For buildings, participation in reserves can be challenging; for example, parameters like ramp-up rate, traditionally defined for generators, are harder to compute for HVAC systems \cite{ma_demand_2013}. 

These diverse DR services are further characterized by their specific operational timing parameters, as summarized in Table \ref{demand_response_timing}, which details typical procurement windows, ramp periods, and sustained durations for the service categories previously discussed. Given the characteristics of these DR programs, and our research emphasis on automation and application portability for frequently utilized services, our subsequent analysis will focus on incentive-based Energy, Regulation, and Reserve programs. 

\begin{table}[h!]
\caption{Demand Response Services Timing Parameters \cite{cappers_assessment_2013}}
\label{demand_response_timing}
\vspace{-4mm}
\centering
\scalebox{1}{
\begin{tabular}{p{1.5cm} p{1.8cm} p{1.8cm} p{1.8cm}}
\hline
\textbf{Service Type} & \textbf{Procurement} & \textbf{Ramp \newline Period} & \textbf{Sustained Duration} \\ 
\hline
\textbf{Energy} & Day-ahead or Real-time & Minutes & A few Hours \\
\hline
\textbf{Capacity} & On a yearly basis & Minutes to Hours & Up to 15 Hours \\
\hline
\textbf{Regulation} & Day-ahead or Real-time & 1 Minute to 10 Minutes & Continuous or Varies \\
\hline
\textbf{Operating Reserves} & Day-ahead or Real-time & $\leq$ 10 Minutes & Up to 30 Minutes \\
\hline
\textbf{Secondary Reserves} & Day-ahead or Real-time & 10 Minutes to 30 minutes & 30 Minutes to Hours \\
\hline
\end{tabular}
}
\end{table}

\section{Enrollment \& Qualification}
\label{sec:enrollment}
\subsection{Overview of the process}
Though NAESB \cite{NAESBSGTF2010} considers the enrollment \& qualification phase to start with an enrollment request submission, we will include steps prior to that such as learning the requirements of the DR programs, identification of suitable DR programs for the participant, and training of staff related to the regulatory and practical process for participation. The first question that the customer needs to answer is whether to participate through a CSP or directly through the ISO. This decision is typically based on whether the customer has the necessary staff to manage paperwork, learn about the requirements, and take actions throughout the participation, such as bidding and scheduling. In cases where customers do not have such staff, CSPs can help through their expertise. However, even with the help of CSPs, customers find the enrollment process time-consuming and overwhelming, limiting their participation \cite{shoreh_survey_2016}. Additionally, if the customer alone cannot provide the necessary minimum resource size limitation, participation through a CSP (also called an aggregator) becomes mandatory. CSPs provide valuable support and typically reduce risk for the customer, but they also take a share of the profits, thereby reducing the financial benefits of participation. 

Enrollment and qualification decisions are impacted by resource characteristics and control infrastructure in addition to time-related aspects \cite{kiliccote_characterization_2019}. Although slow response times can be achieved manually, telemetry and automation become mandatory in programs such as regulation \cite{paterakis_overview_2017}. Although this seems like a limitation, in another way, this brings the benefit that customers can participate in DR programs that suit their resources. However, the variation in DR program requirements across ISOs has been identified as a barrier to broader participation in DR \cite{faria_overview_2015, macdonald_demand_2023}. For a detailed comparison of ISOs' different DR programs, a review is available in \cite{helman_demand_2021}. This has also limited the ability of energy management and control system vendors to provide revenue offerings to customers participating in different markets \cite{kiliccote_characterization_2019}. In addition, it has been identified that the reduction amount is difficult to establish when the load variability is high and resources are weather sensitive \cite{PacifiCorp2013}, such as HVAC systems, where load profiles are highly weather-dependent. Further, customers' acceptance of DR programs is often dependent on their acceptance of automation and control technology, such that customers should be able to ``set it and forget it'' \cite{cappers_assessment_2013}. Consequently, the authors in \cite{cappers_assessment_2013} offered unbundling incentives and automation/control technology with DR programs. Alterations to program requirements have also been offered to allow more DR products to be able to participate in the programs. Changes have included modifying the characteristics of performance and enabling infrastructure, reducing their cost, redefining the bulk power system, and increasing the benefits \cite{cappers_assessment_2013}. An example of the system definitions that need to be changed is from the reserves domain. As generators have traditionally provided these services, there are still parameters, such as the ramp-up rate, that are hard to compute for HVAC systems \cite{ma_demand_2013}.

\begin{table}[htb] 
\caption{Defined Information Requirements for the Enrollment \& Qualification Stage, extracted from \cite{NAESB2008}.}
\label{enrollment}
\centering 
\scalebox{1}{ 
\begin{tabular}{l} 
\textbf{Information Requirements} \\ 
\hline
\textbf{Product Features \cite{NAESB2008, NAWEDRPC2018}} \\
- Minimum Eligible Resource Size \\
- Minimum Reduction Amount \\
- Availability \\
- Aggregation Allowed \\
\hline
\textbf{After-The-Fact Metering \cite{NAESB2008, NAWEDRPC2018}} \\
- After-the-Fact Metering \\
- Meter Interval \\
- Meter Accuracy \\
- Meter Data Reporting Deadline \\
\hline
\textbf{Telemetry \cite{NAESB2008, NAWEDRPC2018}} \\
- Telemetry \\
- Communication Protocol \\
- Telemetry Reporting Interval \\
- Telemetry Accuracy \\
\hline
\textbf{Event Timing \cite{NAESB2008, NAWEDRPC2018}} \\
- Advance Notification(s) \\
- Lead Time for Reduction \\
- Sustained Response Period \\
- Recovery Period \\
- Non-Participation Notice \\
\hline
\end{tabular}
} 

\end{table}

\subsection{Identification of Information Requirements}
The attempt to establish DR standards aimed at unifying the industry and enhancing the liquidity of demand products in wholesale electricity markets has been unsuccessful \cite{coe_demanding_2010}. The authors in \cite{coe_demanding_2010} draw a comparison between the process of transitioning to the euro currency in Europe and the standardization of DR programs. However, despite these efforts, a unified standard remains elusive due to numerous barriers, one of which is inevitably politics. Variations among ISO DR programs hinder commercial building participation, as building operators must undertake an extensive process to (1) identify the qualification requirements and (2) check whether they can satisfy those requirements. An effort against market variety was carried out by the ISO / RTO council in 2018 \cite{NAWEDRPC2018} by collecting the features of each DR program for each ISO in North America. However, this information has not been updated since then. To assess how much has changed over five years, we reviewed and updated PJM’s DR program information. We found that many programs have been replaced, and certain requirements have changed within existing programs. Some efforts have been made to decouple control technologies from DR programs—for example, through the introduction of smart thermostats—though these have primarily targeted the residential sector. The industry has also attempted to relax participation requirements by reducing minimum resource size thresholds. However, as mentioned above, many definitions are still oriented towards traditional generators and thus limit the participation of DR products. 

Prior work has explored the use of semantic technologies to determine whether a building possesses the appropriate system types and the necessary entities to validate its potential for participation \cite{paul2023open}. In Table \ref{enrollment}, we see IRs that must be obtained to evaluate whether a building or a group of buildings is eligible to participate in a certain DR program. These IRs are a subset of the definitions of NAESB \cite{NAESB2008}, which was also used in \cite{NAWEDRPC2018}. The complete list can be found in \cite{NAESB2008}.

\subsection{Synthesis}
Table \ref{enrollment} provides insights into the requirements during the enrollment \& qualification stage, revealing that this phase predominantly involves dealing with regulatory constraints such as deadlines and the allowance of aggregation. An interesting aspect to note is that while some regulatory constraints, like the existence of telemetry or after-the-fact metering, can be easily identified with the help of an ontology, more nuanced IRs, such as estimating the potential reduction amount a building can offer, require sophisticated data analytics for accurate assessment. Unfortunately, the existing literature offers limited insights into estimating a building's reduction potential prior to participation, a topic that will be explored further in Section \ref{sec:scheduling}. In addition to regulatory constraints, there are technical details to consider, such as the parameters of HVAC systems including meter intervals and telemetry accuracy. Interestingly, the enrollment \& qualification stage does not primarily struggle with the volume of data that needs to be communicated to the ISO; rather, the challenge lies in identifying which DR programs each building is eligible for. 

It becomes evident that these IRs would generally fall outside the scope of an upper ontology like Brick, and likely most other building-level ontologies, as they are predominantly policy-based and not directly related to building equipment capabilities. Gathering all these policy requirements for different DR programs, even within the same ISO, is an extremely burdensome task. Consequently, ISOs such as PJM often organize staff training workshops to clarify these requirements, though participation in these workshops typically provides only a general understanding and not a detailed list of the requirements.

\begin{figure}[h!]
\centering
\includegraphics[width=\columnwidth]{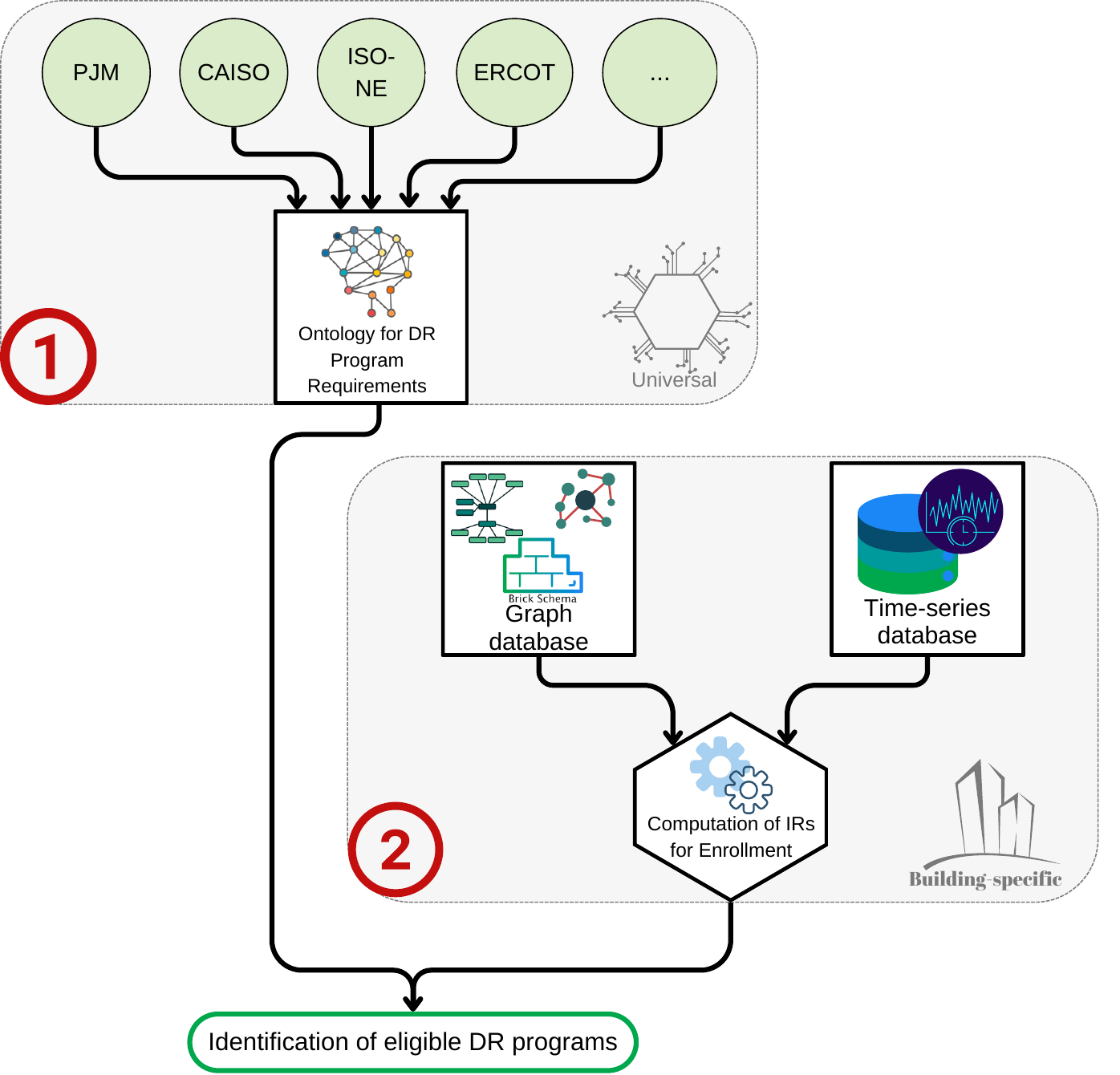}
\caption{Offered Framework for Automation of Enrollment \& Qualification Process}
\label{fig:enrollment}
\end{figure}

As a solution, we offer a future trajectory for the decision-making process in the enrollment \& qualification phase, visually outlined in Figure \ref{fig:enrollment} as an integrated framework showing an information flow from program sources to eligibility assessment. This new structure is aimed at practically removing some of the barriers. As a first step, depicted at the top of the figure, the development of an ontology is needed, which would consolidate inputs from various ISOs. This ontology should have sufficient entities and relationships to represent these policy requirements, as well as technical capabilities. Having this ontology instantiated as a knowledge graph on an online platform, supported by a committee from ISOs and main CSPs, would provide a single reference point for program requirements, reducing barriers related to market variety. Following this, Figure \ref{fig:enrollment} illustrates a subsequent stage where this centralized program information is combined with building-specific data—derived from sources such as a graph database for asset details (potentially leveraging Brick schema) and a time-series database for operational data. This combined information then feeds into a computational process designed to evaluate a building's IRs against the program criteria. While existing schemas like Brick are valuable for asset representation, ensuring all necessary building-specific IRs are available for this computational comparison may require extensions, as discussed later in Section \ref{sec:roadmap}. However, certain IRs, such as minimum eligible resource size or reduction amount, remain challenging to compute accurately even within such a framework. Thus, future research should collaborate in establishing predictive methodologies to estimate resource size for this computational step. This also helps to remove the middle manager (i.e., aggregators) in the case of buildings satisfying minimum requirements. Lastly, the framework culminates, as shown in the figure, in the identification of eligible DR programs; this platform, similar to Enel North America's calculator \cite{enel_demand_response_calculator}, should also allow users to assess expected annual revenue and necessary infrastructure costs for participation.

\section{Scheduling \& Award Notification}
\label{sec:scheduling}
\subsection{Overview of the process}
Scheduling \& award notification can be decomposed into four stages: prediction of the energy reduction potential, bidding into the market, award notification, and scheduling the DR event. Mispredicting the day ahead/intra-day flexibility potential might result in significant penalties or even disqualification from the service. This is also considered one of the main reasons customers hesitate to participate. Thus, scheduling \& award notification is perhaps the most important step of DR participation as it involves the highest uncertainty and directly affects profits. While some studies have explored bidding frameworks for day-ahead energy markets by making use of on-site energy generation and storage systems \cite{parvania_isos_2014}, our primary focus in this stage will be predicting flexibility potential, due to its significant impact on financial and regulatory outcomes of participation. It is worth noting that  scheduling \& award notification are typically required in reserves and energy markets as regulation is mostly automated with the reduction amount being predetermined. Thus, the discussion included in this section will be focused on the scheduling \& award notification process required in energy and reserves services.

\subsection{Identification of the Information Requirements}
We identify three primary data-driven methods that generalize state-of-the-art prediction approaches. First, outside air temperature (OAT)-based prediction models have been developed by simulation studies \cite{yin2016quantifying} as well as experiments \cite{liu2022developing}. A key limitation of OAT-based models is their omission of HVAC system status \cite{de2023unlocking}. Although predictions might state that there is a certain amount of flexibility potential due to a Global Temperature Adjustment (GTA) action, this might be far from real if the VAV is working at the level of minimum ventilation requirements \cite{goddard2014model, de2023unlocking, kiliccote_characterization_2019}. Such errors reduce predictive accuracy and might result in disqualification from the service due to strict requirements for grid services (in the levels of 95\% for reserves \cite{ERCOTProtocols}). Secondly, studies have tried to include the HVAC system dynamics by using damper positions \cite{goddard2014model, vindel2023alphashed}. AlphaShed has shown success in simulation and real-world studies \cite{vindel2023alphashed}. However, AlphaShed predicts instantaneous shedding potential, which limits its applicability for day-ahead or same-day energy reduction predictions. Thirdly, we have DR-Advisor, which is a strategy that uses real-time data and weather forecasts to make predictions \cite{behl2016data}. It mainly differs from prior methods as it tries to assess the effect of control actions at the time of the event and optimize for the actions to be taken to achieve the promised reduction. By doing so, it moves the challenge from the prediction phase to the control phase. One limitation of this approach is that it assumes the amount of reduction that needs to be satisfied is known. While this may hold for participants with fixed reduction agreements with CSPs, participants who work with ISOs directly need to decide this amount themselves.

\begin{table}[h!]
\caption{Defined Information Requirements for the Scheduling \& Award Notification Stage}
\label{tab:forward}
\centering
\scalebox{1}{ 
\begin{tabular}{l}
\textbf{Information Requirements} \\ 
\hline
\textbf{AlphaShed Training \cite{vindel2023alphashed}} \\
- VAV damper positions \\
- Minimum VAV damper positions \\
- Building electricity consumption \\
- Building baseline load predictions \\
- HVAC electricity consumption \\
- HVAC baseline load predictions \\
- The terminal box size \\
- Supply airflow setpoint \\
- HVAC system design size \\
- Supply fan airflow rate \\
- Past DR schedule \\
\hline
\textbf{AlphaShed Deployment \cite{vindel2023alphashed}} \\
- VAV damper positions \\
- Zone air temperature setpoint \\
- Future DR schedule \\
- Occupancy schedule \\
\hline
\textbf{OAT-based Training \cite{yin2016quantifying}} \\
- Outside air temperature \\
- Outside air temperature breakpoints \\
- Building electricity consumption \\
- Building baseline load predictions \\
- Past DR schedule \\
\hline
\textbf{OAT-based Deployment \cite{yin2016quantifying}} \\
- Outside air temperature forecast \\
- Building baseline load predictions \\
- Future DR schedule \\
\hline
\textbf{DR-Advisor Training \cite{behl2016data}} \\
- Outside air temperature \\
- Relative humidity \\
- Wind characteristics \\
- Global horizontal irradiance \\
- Day of the week \\
- Time of day \\
- Chilled water supply temperature setpoint \\
- Hot water supply temperature setpoint \\
- Zone air temperature setpoint \\
- Supply air temperature setpoint \\
- Lighting levels \\
\hline
\end{tabular}
} 

\end{table}
\subsection{Synthesis}
Table \ref{tab:forward} delineates IRs from a variety of high-level categories, illustrating the substantial need for detailed building operation settings and measurements such as setpoints and damper positions. These are closely followed by the necessity for environmental factors and their forecasts, which are crucial for dynamic building management. Additionally, all use cases consistently require timing-based information related to DR scheduling and management, highlighting the need for entities regarding the scheduling of DR events. Other essential IR categories include forecasts of baselines and parameters of HVAC systems, such as terminal box sizes. Energy consumption and metering data, particularly regarding building or HVAC electricity usage, are also identified as critical IRs. Collectively, these requirements show that the Scheduling \& Award Notification stage encompasses a broad spectrum of IRs, reflecting the complexity and multifaceted nature of this phase.

It is important to note that our analysis focuses primarily on methodologies presented in existing literature, which may not be widely adopted in practice. This focus is partly due to the lack of comprehensive studies surveying current practices in commercial buildings, which significantly limits our understanding of the practical applications of these methodologies. One study found that participants often struggle to deliver committed energy curtailments \cite{kang2018data}; but lacks detailed insight into how these reductions are predicted in advance. In one instance, a participant’s energy consumption increased during a DR event, underscoring the critical need for accurate predictions of flexibility potential. Although the reasons why these methods do not transition beyond the pilot stage remain unclear, reducing the engineering efforts required for deploying these applications by allowing semantic interoperability could potentially lower some of the barriers to broader implementation.

\section{Deployment \& Real-Time Communications}
\label{sec:deployment}
\subsection{Overview of the process}
\begin{figure}[h]
\centering
\includegraphics[width=1\columnwidth]{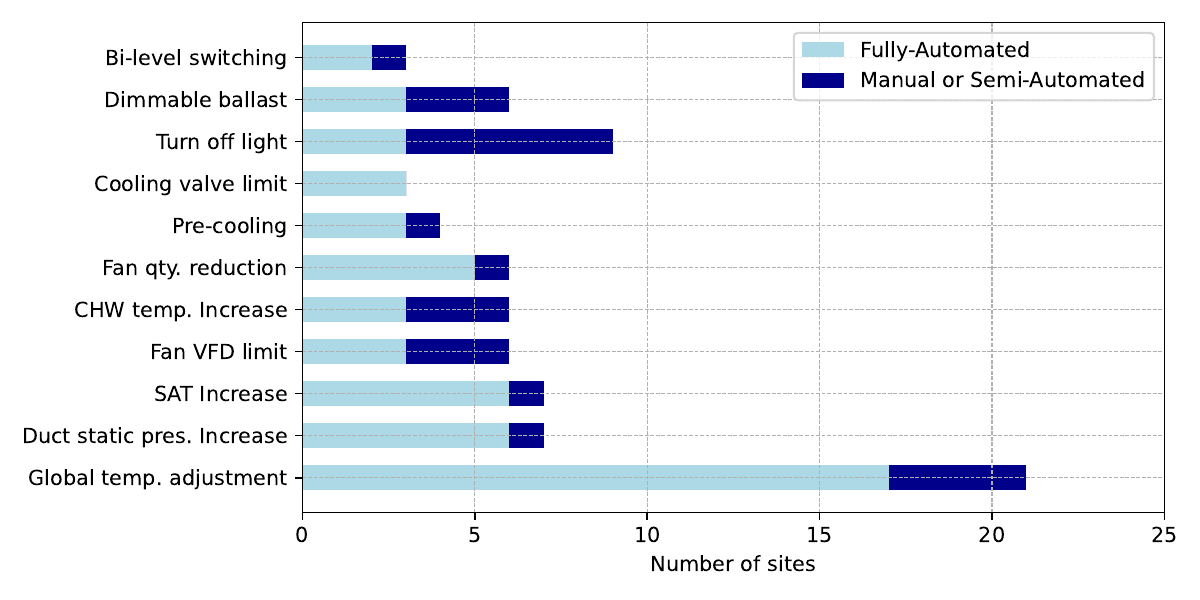}
\caption{DR strategies used by 28 commercial facilities in 2006 (reproduced from \cite{watson2006strategies} with permission).}
\label{fig:strategies}
\end{figure}

If the bid is approved, necessary actions should be taken in the given DR period to achieve the promised reduction. How these actions are implemented varies significantly across applications. Since control configurations and reduction strategies are the customer’s responsibility, ISOs do not have any specifications related to the strategies that can be taken. A study published in 2006 analyzed the DR strategies used in 28 commercial buildings through field tests \cite{watson2006strategies}. The study examined the distribution of these strategies and their level of automation, as shown in Figure \ref{fig:strategies}. The results revealed a considerable level of automation, with GTA being the most frequently used strategy. However, it is important to note that the high level of automation observed in these facilities may be due to the study being a pilot. Consequently, it remains unclear how often these actions are used in practice and what the current level of automation is. Moreover, since the study was conducted almost 20 years ago, its applicability today is limited. Therefore, surveys are necessary to determine how customers are providing the required reduction. Some surveys have been conducted to examine the practice in industrial applications \cite{samad2012smart, shoreh_survey_2016}. In certain situations, such as in the case of big refrigeration plants, turning off the equipment through automatic switches for a few minutes (i.e. regulation) won't cause any disturbance in the service. However, for other processes like cement manufacturing, energy-intensive products such as grinding mills can be scheduled to stop working and use the stored material during energy dispatching events. However, their unique processes are not applicable to commercial buildings. Thus, we draw on our experience with facility managers to describe current practices in commercial buildings. In the following paragraphs, we will go over regulation, energy, and reserves applications in commercial buildings and extract the IRs based on the information used in each study.

\subsection{Identification of the Information Requirements}
Regulation services require fast response times, making automation essential. In \cite{beil_frequency_2016}, the authors showed that zone-based setpoint adjustments result in considerable latency compared to the standards of ISOs and thus limit participation in regulation. However, fan power or supply pressure control achieves the necessary response time but cannot manage to condition occupied and unoccupied zones differently. Also, batteries and water heaters are stated as resources for regulation services with the necessary communication protocol. 

Energy dispatching, whether real-time or day-ahead can be conducted manually due to its longer notification time and larger reduction requirement. One common misconception exists in the literature where real-time energy dispatching is sometimes assumed to be truly ``real-time''. However, in reality, it refers to biddings available intra-day (usually a couple of hours before the event). If participating in energy dispatching through a CSP, the bidding is conducted on their side and the participant is tasked with providing the reduction required by the CSP. In these cases, a typical day of participation would be as follows: a signal (email, phone call, or message) is sent by the aggregator to the facility management offices of big customers (e.g., universities). After receiving the signal, facility personnel may search for unoccupied zones (for the DR period ahead of time) to increase the deadband of the setpoint, also called GTA \cite{behl2016data}. In conditions where only focusing on unoccupied zones would not produce enough reductions, they may increase the deadbands in occupied zones too. Depending on ISO regulations, they can precool or preheat these zones to reduce the discomfort and/or increase the duration of the DR event. 

Reserve services include synchronized, non-synchronized, and secondary (operating) reserves. Synchronized and non-synchronized reserves should be ramping up in less than 10 minutes while the former needs to be automated. Non-synchronized reserves are expected to provide the same performance without automation so they can be semi-automated or manual. Secondary reserves, on the other hand, can ramp up between 10 to 30 minutes and thus they are either semi-automated or manual. Although buildings have traditionally been used in day-ahead or real-time energy markets with notifications varying from 2 to 24 hours, A study by Kiliccote et al. \cite{kiliccote_characterization_2019} showed the potential in buildings to participate in price-based non-spinning reserves. DR is still not permitted in non-spinning reserves by certain major ISOs (e.g., PJM \cite{PJM2024Manual11}). However, we see that commercial buildings can satisfy the 10-minute response requirement using OpenADR. Their methodology mapped four price signals to 4°F setpoint changes with 1°F increments. Ramp rates were significantly higher than predictions because the fans immediately went to minimum performance mode. Their algorithm also used constant shedding predictions with 4-second telemetry. This application was successful but required significant engineering effort to deploy and match all the algorithms with the necessary control points. Another study showed that a hotel can participate in spinning reserves with responses occurring between 12 to 60 seconds \cite{kirby2008spinning}. DR-Advisor is another method that can be used in reserve markets due to its ability to evaluate DR potential in real-time. It considers the weather forecast and the current state of the building to investigate the most suitable DR strategy in real time and executes it. This is especially useful in complex building systems where it is hard to understand the effect of taking an action due to dependencies on other subsystems. Moving the uncertainty from prediction to control with a hierarchy of control options available would reduce the overall risk of penalties or disqualification from the service. 

EV charging infrastructure has started to become a part of commercial buildings and it provides significant opportunities for DR \cite{pourghaderi_commercial_2018}. Participation in DR with EV charging can be through, energy, reserves, or regulation \cite{brooks_demand_2010} as it has considerable flexibility and potential for automation. Thus, it is considered separately. To extract the IRs for EV charging, we focused on a combination of the following works. In \cite{calearo2021review}, the authors conducted a review of the data sources required for EV integration and provided the variables required in an ideal dataset. Zhao et al. created a class diagram for object communications including PHEVs and Battery storage in a residential house \cite{zhao_integrated_2013}. Pourghaderi et al. developed a technical virtual power plant-based bidding framework for the participation of commercial buildings in the day-ahead energy markets, in which HVAC systems and EV charging are considered as DR resources \cite{pourghaderi_commercial_2018}. 

\begin{table}[t!] 
\caption{Defined Information Requirements for the Deployment \& Real-time Communications Stage} 
\label{tab:control} 
\centering 
\scalebox{1}{ 
\begin{tabular}{l} 
\textbf{Information Requirements} \\ 
\hline
\textbf{Regulation \cite{beil_frequency_2016}} \\
- Fan power (VFD) \\
- Supply pressure setpoint \\
- DR start and end signal \\
\hline
\textbf{Global Temperature Adjustment (GTA) \cite{behl2016data}} \\
- Occupancy schedule \\
- Zone air temperature deadband \\
- Zone air temperature setpoint \\
- Future DR schedule \\
- Preheating/cooling allowance \\
- Preheating/cooling start time \\
\hline
\textbf{Price Based Non-Spinning Reserves \cite{kiliccote_characterization_2019}} \\
- Zone air temperature setpoint \\
- Zone air temperature deadband \\
- Future DR schedule \\
- DR start and end signal \\
- Price level \\
- Price - setpoint adjustment mapping \\
\hline
\textbf{EV charging station \cite{zhao_integrated_2013, pourghaderi_commercial_2018, calearo2021review}} \\
- State of charge \\
- Battery capacity \\
- Maximum and minimum charging voltage \\
- Discharging allowance \\
- Arrival time \\
- Departure time \\
- Charging voltage \\
- Charging current \\
- EV energy consumption \\
- Charging status \\
- Charged energy \\
- Charging power \\
\hline
\textbf{DR-Advisor Deployment \cite{behl2016data}} \\
- Outside air temperature forecast \\
- Relative humidity forecast \\
- Wind characteristics forecast \\
- Solar irradiation forecast \\
- Day of the week \\
- Time of day \\
- Chilled water supply temperature setpoint \\
- Hot water supply temperature setpoint \\
- Zone air temperature setpoint \\
- Supply air temperature setpoint \\
- Lighting levels \\
\hline
\end{tabular}
} 

\end{table}

\subsection{Synthesis}
In examining the application areas detailed in Table \ref{tab:control}, it is clear that the Deployment \& Real-time Communications stage shares many IRs with the Scheduling \& Award Notification stage. This overlap is largely due to the fact that the control actions implemented during this stage are based on predictions made in the previous stage. Predominantly, the IRs required again revolve around building operation settings, highlighting the continuous need for managing and monitoring these settings effectively. Additionally, environmental factors and their forecasts play a crucial role, particularly in applications like DR-Advisor, where accurate environmental data can significantly impact decision-making processes. In the context of EV charging, unique IRs arise from the specific control actions associated with this technology. These specific requirements underscore the specialized nature of EV charging systems and the needed entities for their seamless integration into broader energy management strategies. Furthermore, the concept of DR scheduling and management is critical, particularly in signaling the start and end of DR events and for planning future schedules. This requirement underscores the importance of precise and anticipatory scheduling, which is critical for effective energy management and operational efficiency in real-time energy deployment scenarios.

\section{Measurement \& Performance Evaluation}
\label{sec:measurement}
\subsection{Overview of the process}
Measurement \& Performance Evaluation is a significant step in the DR process since it defines the performance of the participant and thus determines the financial reward \cite{behl2016data}. For example, for energy markets, PJM requires the reduction to be within $\pm$20\% of the bid value to qualify for benefits, meaning that reducing more than needed can also make participants lose profit. Thus, participants are interested in the measurement \& performance evaluation process not only for reporting, but also for obtaining feedback that would increase the success of further operations. While it is common for the ISO to compute the Customer Baseline Load (CBL), participants can recommend alternative methods from among those allowed by the ISO to maximize their profits or minimize the risk of disqualification \cite{PJM2024Manual11}. The CBL models can be grouped into five as shown in Table \ref{table:cbl-methods}. This section focuses on Baseline Types 1 and 2 as they are commonly adopted by the ISOs and have potential for further improvement. 

\begin{table*}[htb]
\caption{Customer Baseline Load Methods \cite{coe_demanding_2010}}
\label{table:cbl-methods}
\centering
\scalebox{1}{ 
\begin{tabular}{p{3cm} p{12cm}} 
\textbf{Method} & \textbf{Description} \\ 
\hline
\textbf{Maximum Base Load} & A method that assesses performance by focusing solely on whether a demand resource can reduce electricity demand to a predetermined level, irrespective of its consumption or demand at the time of deployment. \\
\hline
\textbf{Meter Before/} \\ \textbf{Meter After} & This method compares electricity usage or demand recorded over a set period before deployment with the data collected during the actual response period to evaluate performance. \\
\hline
\textbf{Baseline Type 1} & An evaluation approach that uses historical interval meter data from a demand resource, which may also incorporate other factors like weather conditions and calendar data. \\
\hline
\textbf{Baseline Type 2} & A method that employs statistical sampling to estimate the electricity usage of a group of demand resources, particularly when interval metering is not available for the entire group. \\
\hline
\textbf{Metering Generator Output} & This methodology is applied when a generation asset is behind the demand resource’s revenue meter, with performance measured by the output of that generation asset. \\
\hline
\end{tabular}
} 

\end{table*}

It is important to explain the meaning behind the commonly used terminologies to clarify how the studies were selected for the review. ISOs define the ``counterfactual'' load as the CBL while many studies use energy consumption prediction. There are a few reasons behind these differences. A customer can have multiple buildings that participate in the DR program (e.g., university campuses). Therefore, the ISO is interested in their collective reduction amount which is why they are interested in predicting the Baseline load of their customer rather than a single building. In technical studies, these predictions are investigated at three different scales: appliance level (submetering), building level, or aggregate level \cite{pinto2022towards}. That is why, it is called building baseline load or building energy consumption prediction when the focus is on a single building. While we usually have the lowest error in aggregate level, building level is still needed in cases where the aggregate includes different participants as we need fair financial compensations. 

\subsection{Identification of Information Requirements}

The literature on data-driven prediction of building energy consumption is rich and varied, ranging from simple regression-based methods to advanced deep learning approaches such as the use of artificial neural networks (ANN) dating back to 2000 \cite{kalogirou2000artificial}. A more recent comparative analysis in \cite{pinto2022towards} evaluated four different models: a multi-layer perceptron feed-forward neural network, extreme gradient boosting, OAT-based piece-wise linear regression, and Mid4of6. Interestingly, while the latter model is industry-recognized, the former three, particularly the OAT-based and Extreme Gradient Boosting models, were found to be superior in performance. In a creative crossover of disciplines, Xue and Salim (2023) explored how natural language processing models—Bart, Bigbird, and Pegasus—could be applied to energy load forecasting \cite{xue2023utilizing}. Additionally, a study within the DR-Advisor project introduced a tailored family of regression trees for CBL prediction \cite{behl2016data}. For those seeking a deeper dive into these methodologies, detailed reviews are available \cite{amasyali2018review, zhang2021review}. It is also noteworthy to mention that while not central to our discussion, most ISOs typically calculate compensation based on five-minute intervals, derived from aggregated hourly metering data, highlighting the precision required in contemporary energy management practices.

On the other hand, existing practices have been using mostly averaging-based techniques. The most commonly known one is HighXofY which uses Y days similar to the event day and takes the highest X of them for averaging \cite{johra2023iea}. In addition, PJM offers Weather Sensitive Adjustment (WSA) to its CBLs to account for an increase or decrease in consumption due to weather-related events \cite{PJM2024Manual11}. It is important to note that WSA does not predict the CBL but rather it adjusts the CBL's predictions using weather. Participants can also recommend other CBL prediction methods and, if approved by the ISO, they may be used. Similarly, ISOs can request changes in the CBL computation methods and are considered accepted if the participants do not reply within a certain period. In the case of a single building, this process may seem easy and not require a semantic ontology to extract the necessary metering data. However, in commercial building facilities such as universities where multiple buildings may participate in a DR event, it may be difficult to identify and extract the energy consumption data of each building. Additionally, in the case of achieving full autonomy in DR, we should be capable of handling data at the district level, which further justifies the need for modeling buildings semantically. Additionally, ASHRAE announced a challenge back in 1994 to explore the simple yet effective methods to predict the energy consumption of buildings as well as chilled and hot water consumption \cite{kreider1994predicting}. Among the papers investigated here, only DR-Advisor utilized this dataset for performance comparison. 

In the realm of energy consumption prediction, existing practices rely predominantly on averaging-based techniques, such as the widely recognized HighXofY method, which averages the highest X days from Y days similar to the day of the event \cite{johra2023iea}. Furthermore, PJM employs a WSA for its CBL calculations, which adjusts for variations in consumption due to weather, rather than predicting new CBL values \cite{PJM2024Manual11}. This framework allows participants to suggest alternative CBL prediction methods, which can be implemented upon approval from ISOs. ISOs also have the authority to propose modifications to CBL computation methods, which are automatically accepted if participants do not respond within a specified period. While the process of extracting metering data might seem straightforward for a single building, it becomes complex in environments like university campuses where multiple buildings might participate in DR programs. This complexity is further compounded when considering data at the district level, highlighting the need for a semantic model of building data. 

\subsection{Synthesis}

\begin{table}[h!]
\caption{Defined Information Requirements for the Measurement and Performance Evaluation Stage}\label{tab:verification}
\centering
\scalebox{1}{ 
\begin{tabular}{l}
\textbf{Information Requirements} \\ 
\hline
\textbf{HighXofY \cite{johra2023iea, PJM2024Manual11}} \\
- X and Y values \\
- Day of the week \\
- Time of the day \\
- Building electricity consumption \\
- Past DR schedule \\
\hline
\textbf{Weather Sensitive Adjustment \cite{PJM2024Manual11}} \\
- Outside Air Temperature \\
- WSA factor \\
\hline
\textbf{DR-Advisor \cite{behl2016data}} \\
- Outside air temperature \\
- Relative humidity \\
- Wind characteristics \\
- Global horizontal irradiance \\
- Day of the week \\
- Time of the day \\
- Past DR schedule \\
- Building electricity consumption  \\
\hline
\textbf{Language Model \cite{xue2023utilizing}} \\
- Day of the week \\
- Time of the day \\
- Building electricity consumption \\
- Past DR schedule \\
\hline
\textbf{ANN \& Extreme Gradient Boosting \cite{pinto2022towards}} \\
- Day of the week \\
- Time of the day \\
- Outside air temperature \\
- Relative humidity \\
- Global horizontal irradiance \\
- Direct normal irradiance \\
- Past DR schedule \\
- Building electricity consumption  \\
\hline
\end{tabular}
} 

\end{table}

Table \ref{tab:verification} illustrates that the measurement and performance evaluation stage primarily requires IRs related to environmental factors, such as outdoor air temperature and global horizontal irradiance, as these factors significantly influence the energy consumption of HVAC systems. Additionally, as expected, data on energy consumption and metering play a critical role as they are frequently used for basing predictions or training models. A notable challenge identified is the extraction of past event data, emphasizing the need for a Past DR Schedule to automatically remove the event days from the pool of available training days, thus underscoring the importance of DR scheduling and management-related IRs. This stage generally involves fewer IR categories compared to earlier stages.

While extracting IRs for a single methodology might not be exceedingly difficult for a single building, Xue and Salim have shown that \cite{xue2023utilizing} despite the superior performance of language models over other complex neural network-based models, no single model adequately met the needs of all six buildings studied. This highlights the different energy consumption patterns of different buildings and supports the need for multiple models, allowing the selection of the most suitable one for each specific case.

However, the requirement to develop multiple models for a single building before identifying one with sufficient accuracy complicates matters. This need for extensive model development, which has not been widely adopted in practice, can be attributed to the substantial engineering effort required and the limited portability of these models across different buildings. Therefore, semantically modeling these IRs could greatly reduce the effort needed for portability, facilitating the broader application of advanced predictive technologies in building energy management.

\section{Categorization of identified IRs}

This section presents the comprehensive inventory of IRs extracted from our review of ISO manuals, NAESB standards, and academic literature. To facilitate the semantic gap analysis, we have organized these IRs into distinct categories based on their functional domain, ranging from static regulatory constraints to dynamic operational settings. Table \ref{table:ir_class} details these categories, listing specific data elements alongside the primary sources that motivate their necessity for automated Demand Response participation.

\begin{table*}[t!]
\caption{Categorization of Information Requirements for Energy Flexibility in Buildings}
\vspace{3mm}
\label{table:ir_class}
\centering
\scalebox{0.9}{ 
\begin{tabular}{|p{6cm}|p{6cm}|p{6cm}|}
\hline
\textbf{Regulatory Constraints} & \textbf{HVAC System Parameters} & \textbf{Building Operation Settings \& Measurements} \\
\hline
Minimum Eligible Resource Size \cite{NAESB2008, NAWEDRPC2018} & Minimum VAV damper positions \cite{vindel2023alphashed} & Zone air temperature deadband \cite{behl2016data, kiliccote_characterization_2019} \\
Minimum Reduction Amount \cite{NAESB2008, NAWEDRPC2018} & Terminal box size \cite{vindel2023alphashed} & Zone air temperature setpoint \cite{vindel2023alphashed, behl2016data, kiliccote_characterization_2019} \\
Availability \cite{NAESB2008, NAWEDRPC2018} & HVAC system design size \cite{vindel2023alphashed} & Chilled water supply temperature setpoint \cite{behl2016data} \\
Aggregation Allowed \cite{NAESB2008, NAWEDRPC2018} & Meter Interval \cite{NAESB2008, NAWEDRPC2018} & VAV damper positions \cite{vindel2023alphashed} \\
After-the-Fact Metering \cite{NAESB2008, NAWEDRPC2018} & Meter Accuracy \cite{NAESB2008, NAWEDRPC2018} & Hot water supply temperature setpoint \cite{behl2016data} \\
Meter Data Reporting Deadline \cite{NAESB2008, NAWEDRPC2018} & Telemetry Reporting Interval \cite{NAESB2008, NAWEDRPC2018} & Supply air temperature setpoint \cite{behl2016data} \\
Telemetry \cite{NAESB2008, NAWEDRPC2018} & Telemetry Accuracy \cite{NAESB2008, NAWEDRPC2018} & Supply airflow setpoint \cite{vindel2023alphashed} \\
Communication Protocol \cite{NAESB2008, NAWEDRPC2018} & & Supply fan airflow rate \cite{vindel2023alphashed} \\
Advance Notification(s) \cite{NAESB2008, NAWEDRPC2018} & & Lighting levels \cite{behl2016data} \\
Lead Time for Reduction \cite{NAESB2008, NAWEDRPC2018} & & Fan power (VFD) \cite{beil_frequency_2016} \\
Sustained Response Period \cite{NAESB2008, NAWEDRPC2018} & & Supply pressure setpoint \cite{beil_frequency_2016} \\
Recovery Period \cite{NAESB2008, NAWEDRPC2018} & & Price - setpoint adjustment mapping \cite{kiliccote_characterization_2019} \\
Non-Participation Notice \cite{NAESB2008, NAWEDRPC2018} & & Occupancy Schedule \cite{vindel2023alphashed, behl2016data} \\
\hline

\textbf{Demand Response Scheduling and Management} & \textbf{Electric Vehicle (EV) Charging Infrastructure} & \textbf{Environmental Factors and Forecasts} \\
\hline
Past DR schedule \cite{vindel2023alphashed, yin2016quantifying, johra2023iea, xue2023utilizing, PJM2024Manual11, pinto2022towards} & State of charge \cite{zhao_integrated_2013, pourghaderi_commercial_2018, calearo2021review} & Outside air temperature \cite{yin2016quantifying, behl2016data, PJM2024Manual11, pinto2022towards} \\
Future DR schedule \cite{vindel2023alphashed, yin2016quantifying, behl2016data, kiliccote_characterization_2019} & Battery capacity \cite{zhao_integrated_2013, pourghaderi_commercial_2018, calearo2021review} & Relative humidity \cite{behl2016data, pinto2022towards} \\
DR start and end signal \cite{beil_frequency_2016, kiliccote_characterization_2019} & Maximum and minimum charging voltage \cite{pourghaderi_commercial_2018} & Wind characteristics \cite{behl2016data} \\
Preheating/cooling allowance \cite{behl2016data} & Discharging allowance \cite{zhao_integrated_2013} & Global horizontal irradiance \cite{behl2016data, pinto2022towards} \\
Preheating/cooling start time \cite{behl2016data} & Arrival time \cite{pourghaderi_commercial_2018, calearo2021review} & Direct normal irradiance \cite{pinto2022towards} \\
Price level \cite{kiliccote_characterization_2019} & Departure time \cite{pourghaderi_commercial_2018, calearo2021review} & Outside air temperature forecast \cite{yin2016quantifying, behl2016data} \\
X and Y values \cite{johra2023iea, PJM2024Manual11} & Charging voltage \cite{zhao_integrated_2013} & Relative humidity forecast \cite{behl2016data} \\
& Charging current \cite{zhao_integrated_2013} & Wind characteristics forecast \cite{behl2016data} \\
& EV energy consumption \cite{zhao_integrated_2013, pourghaderi_commercial_2018} & Solar irradiation forecast \cite{behl2016data} \\
& Charging status \cite{zhao_integrated_2013} & \\
& Charged energy \cite{calearo2021review} & \\
& Charging power \cite{calearo2021review} & \\
\hline

\textbf{Time-Based Parameters} & \textbf{Forecasts of Energy Baseline} & \textbf{Energy Consumption and Metering} \\
\hline
Day of the week \cite{behl2016data, johra2023iea, xue2023utilizing, PJM2024Manual11, pinto2022towards} & Building baseline load predictions \cite{vindel2023alphashed, yin2016quantifying} & Building electricity consumption \cite{vindel2023alphashed, yin2016quantifying, johra2023iea, xue2023utilizing, PJM2024Manual11, pinto2022towards} \\
Time of the day \cite{behl2016data, johra2023iea, xue2023utilizing, PJM2024Manual11, pinto2022towards} & HVAC baseline load predictions \cite{vindel2023alphashed} & HVAC electricity consumption \cite{vindel2023alphashed, yin2016quantifying, xue2023utilizing} \\
& OAT breakpoints \cite{yin2016quantifying} & \\
\hline

\end{tabular}
} 

\end{table*}

\section{Required Ontology Extensions}\label{sec:roadmap}
Based on the gaps identified in the previous sections, this section proposes specific extensions to the Brick schema and related ontologies to fully support the automation of DR processes.

\subsection{Regulatory constraints}

In Section \ref{sec:enrollment}, we highlighted the need for (1) established computation strategies for challenging-to-obtain IRs, such as minimum reduction amount and minimum eligible resource size, (2) a centralized ontology or framework that consolidates the requirements for each DR program across ISOs, and (3) concepts capable of modeling or inferring whether buildings can meet these IRs, including aspects like communication protocols or telemetry types.

We proposed a future direction for the decision-making process during the enrollment phase, as illustrated in Figure \ref{fig:enrollment}. This proposed structure aims to streamline decision-making and effectively remove some existing barriers. The first step involves creating a new centralized ontology or extending established utility standards with adequate entities and relationships to represent both policy requirements and technical capabilities. Hosting a knowledge graph created by using this ontology on an online platform, maintained by a committee of representatives from each ISOs and major CSPs, would provide a unified source of information on program requirements, significantly lowering the barriers posed by market diversity. Certain IRs, such as minimum eligible resource size or reduction amount, remain difficult to calculate. Therefore, further studies must focus on developing predictive methods that can offer a preliminary estimation of resource size. This approach could also reduce the need for intermediaries, such as aggregators, for buildings that meet the minimum requirements. Finally, similar to the tool provided by Enel North America \cite{enel_demand_response_calculator}, this platform should enable users to identify eligible programs, calculate their expected annual revenue, and estimate the infrastructure costs required for participation in more DR programs.

While neither Brick nor the previously discussed DR-focused ontologies currently have the capability to represent these regulatory constraints, the CIM provides a substantial, though incomplete, foundation. Our analysis reveals that CIM (specifically spanning IEC 62325-301, IEC 61968-9, and IEC 61970-301) covers over half of the identified regulatory IRs. It is particularly effective at modeling market participation rules and resource limits, including Minimum Eligible Resource Size (\texttt{ResourceCapacity.minimumCapacity}), Minimum Reduction Amount (\texttt{LoadBid.minLoadReduction}), Lead Time for Reduction (\texttt{LoadBid.reqNoticeTime}), and Sustained Response Period (\texttt{RegisteredResource.minDispatchTime}). Furthermore, it captures sensing hardware information such as Meter Interval
(\texttt{MeasuringPeriodKind}) and networking information such as Communication Protocol (\texttt{ComTechnologyKind}).

However, CIM is not without critical gaps. It entirely lacks entities for Telemetry requirements (existence, reporting intervals, and accuracy), which are strictly enforced by many ISOs. It also fails to capture temporal operational constraints such as Recovery Periods, Advance Notifications, and After-the-Fact Metering deadlines. Furthermore, CIM is not an open-source standard; its proprietary nature under the IEC creates a significant barrier to entry, restricting accessibility for researchers, smaller aggregators, and community-driven development. Therefore, while the market-facing components of CIM serve as a valuable conceptual reference, developing a truly open-source ISO-focused ontology would be far more accessible and practical for widespread adoption. This open framework should be represented by a knowledge graph and integrated with the necessary computational tools, as explained above.

\subsection{HVAC System Parameters}

HVAC System Parameters are fixed parameters that are not linked to any database. Brick includes the property \texttt{value} for assigning such fixed values. Conversely, for parameters related to intervals or accuracy, Brick lacks relations such as \texttt{HasInterval} or \texttt{HasAccuracy}. In contrast, for telemetry modeling, OpenADR Ontology (oadr) provides classes like \texttt{TelemetryReport}, \texttt{TelemetryUsageReport}, and \texttt{Telemetry\allowbreak Status\allowbreak Report}. These classes feature object properties such as \texttt{hasAccuracy} and \texttt{hasSamplingRate}, enabling the modeling of telemetry accuracy and reporting intervals. Similarly, the CIM standard provides established entities for meter intervals (\texttt{MeasuringPeriodKind}) and accuracy (\texttt{MeasurementValue. sensorAccuracy}), although it lacks specific telemetry attributes. Ultimately, meters can be modeled in Brick, and their accuracy and interval settings can be represented through these relationships from DELTA or CIM.

\subsection{Building Operation Settings \& Measurements}

We observe that Brick is successful in modeling almost all of these setpoints and measurements. Although Brick can access occupancy sensors in real-time, in certain cases, future occupancy schedules (i.e., Occupancy Schedule of Zones) is needed to ensure which zones' setpoints can be changed in advance. This is a class that could be included in Brick through extensions. Additionally, for certain applications (e.g., reserves), to achieve automation with price-based DR, we would need price-setpoint matching. While Brick lacks this capability, CIM successfully addresses it using the \texttt{LoadReductionPriceCurve} entity. Since this functionality is not fully related to building operations or DR management, it lies between both domains, and thus it could either be mapped to CIM or modeled with an extension of Brick using CIM as a conceptual reference.

\subsection{Demand Response Scheduling and Management}

Previous studies have also identified that most DR classes were not available in Brick or in ASHRAE 223\cite{paul2023open, de2024enabling}, necessitating self-configuration efforts to provide application portability. This is not unexpected since Brick was not specifically designed for DR purposes. Our aim in this section is to identify the appropriate extensions and integrations. Thus, we also review DR-specific ontologies and utility-scale models like CIM and then explain how they can be used to support Brick.

DELTA uses the OpenADR Ontology class \texttt{Event}, which can be used to describe a DR event. It includes certain relationships such as \texttt{hasDuration} or \texttt{has\allowbreak Delivery\allowbreak Time}, which are useful for modeling upcoming DR events. However, they are not modeled under the same namespaces, which might cause confusion in querying, as they are closely related concepts. Therefore, there should be an extended version where we can explicitly model future DR events and their starting and ending times.

DELTA appears to be successful in modeling the energy market, with entities for bid prices for ancillary services, day-ahead, or intra-day markets. Additionally, it includes an entity for capacity sold, but not one for reduction sold. While the difference between the two may not be immediately apparent, capacity typically refers to the designated capacity that a participant assigns to the grid, for which the participant can be paid even if the event is not called. Reduction, on the other hand, refers to the amount of energy consumption reduction promised within a given period, with penalties applied if not met. Thus, these two elements should be modeled separately.

The OpenADR Ontology also has a class named \texttt{schedule}, which can be used with the property \texttt{isScheduleOf}. Together, these can model the schedule of an event. However, one concern is that this schedule should include the start and end times of a particular event. To create a more unified data storage method, we might need extensions such as \texttt{hasStartingTime} and \texttt{hasEndingTime}.

Having a separate class for past DR schedules is needed for application development, especially when historical data is required. It is worth noting that CIM completely fails to provide functional entities for either past or future DR schedules, reinforcing the necessity of custom extensions. In terms of modeling signals and price levels, both DELTA and OpenADR Ontology are effective. Furthermore, CIM offers robust support for the operational transmission of these event parameters through its IEC 61968-9 standard, explicitly using \texttt{EndDeviceControl.priceSignal} for mapping price levels and \texttt{EndDeviceControl.scheduledInterval} to define DR start and end signals. However, neither the DR-specific ontologies nor CIM has classes to model preheating/cooling allowances or their specific pre-event start and end times.

In EFOnt, the subgroup \texttt{ThermallyActivatedBuildingSystems} includes \texttt{pre\_heating}, \texttt{pre\_cooling}, and \texttt{temperature\_setpoint\_adjustment}. An example provided in EFOnt's repository illustrates that \texttt{pre\_heating} and \texttt{pre\_cooling} have a \texttt{canBeModeledBy} relationship with \texttt{ThermostatSetpoint}, suggesting that an EnergyPlus model can estimate the response of these actions. However, in real-life applications, many buildings lack detailed EnergyPlus models, leading to reliance on simpler rules of thumb. For instance, during our interaction with a local Facility Management System (FMS) team, we learned that preconditioning spaces is typically performed one to two hours before a DR event. While this diverges from the intended precision of EFOnt, the conceptual framework can still be leveraged, albeit with a certain "abuse of ontology use." Specifically, we can represent preconditioning activities using \texttt{pre\_heating} or \texttt{pre\_cooling} from EFOnt. To enhance this representation, we can also utilize previously proposed relationships, such as \texttt{hasStartingTime}, \texttt{hasEndingTime}, and add another one (\texttt{Setpoint\_Change}), to formalize the timing and scope of these preconditioning actions, successfully bridging a critical gap left unaddressed by CIM, Brick, and OpenADR.

\subsection{EV charging Infrastructure}
Brick does not currently have any class for electric vehicles. In EFOnt, there is an \texttt{ElectricVehicle} class, but it only includes a relation on how it can be modeled. Using another class from EFOnt (\texttt{GenericLoadProfile}), its' energy consumption can be modeled. In addition, the battery capacity can be modeled with \texttt{Battery} and \texttt{StorageCapacity} from DELTA. However, while building-centric ontologies struggle with performance metrics such as state of charge, and current and voltage limits, CIM actually excels here. Through its IEC 61970-301 standard, CIM successfully captures these physical and electrical properties using entities like \texttt{Battery Unit.storedE} (state of charge), \texttt{BatteryUnit.ratedE} (capacity), \texttt{VoltageLimit}, and \texttt{CurrentFlow}. 

Despite this electrical precision, CIM still fails to model the behavioral and mobile nature of EVs, lacking entities for arrival and departure times, discharging allowances, and dynamic EV energy consumption. We have identified another ontology specific to EVs, but it focuses on life cycle management \cite{yoo2016semantic}, making it unsuitable for DR-related properties. Given CIM's proprietary restrictions and its blind spots regarding EV mobility, it appears that the simplest and most accessible solution may be to extend Brick itself for this purpose, potentially adopting CIM's electrical entity structures as a conceptual baseline, as EV Charging Infrastructure is already becoming a part of many commercial buildings.

\subsection{Energy Consumption and Metering}
In this context, Brick is quite successful as we do not observe any unsatisfied IRs. However, one area that caught our attention is the modeling of the Coefficient of Performance (COP) in cases where submetering is available. HVAC power can consist of electric power components from an air handling unit, a chilled water pump, a hot water pump, and an exhaust fan. However, there may be elements measured in thermal power units (e.g., chiller thermal power), which would require conversion to electric power to accurately compute the HVAC power. This is where the COP value becomes necessary. Therefore, we might need a class (e.g., \texttt{Coefficient\_Of\_Performance}) to model the COP value (either as a constant with \texttt{value} or as a time series with \texttt{ref:TimeseriesReference}) within Brick.

\subsection{Time Based Parameters}
Time-Based Parameters include values such as Day of the Week or Time of the Day. Since time-series databases store data with timestamps, these IRs can be derived directly from the timestamp, meaning Brick can already satisfy these requirements. The only extension that might be necessary is for cases when certain days are considered off-business days. Off-business days would need to be treated like weekends in office buildings. Treating them as ordinary weekdays could lead to inconsistencies in the data required to predict the energy baseline. This issue can be addressed by storing official holidays in a database and modeling them as an entity, such as \texttt{Off\_Business\_Schedule}.

\subsection{Forecasts of Energy Baseline}
This IR category is different from others because it is neither a measurement nor a fixed value but rather the output of a model. While the CIM standard contains a forecast class for grid and weather forecasts (discussed below), it entirely lacks the semantic entities required to represent building-level or HVAC-level baselines. In Section \ref{sec:measurement}, we discussed that the goal should not be to find a single model that works for all buildings, but rather to develop a library of models from which the best one for each building can be selected. If an established model is available for a particular building, a simple script can automatically make predictions for the next day and store them in a database as if they were measured values. For this, we would need straightforward classes such as \texttt{Building\_Baseline\_Predictions} and \texttt{HVAC\_Baseline\_Predictions}, to which we could assign \newline \texttt{ref:TimeseriesReference} that is already used by Brick. As a secondary strategy, we can introduce a property like \texttt{hasForecast\allowbreak Reference}, drawing conceptual inspiration from CIM's forecast modeling approach. This approach would still require the definition of classes such as \texttt{HVAC\_Baseline} or \texttt{Building\_Baseline} to maintain clarity and structure within the ontology, ensuring that model predictions are properly categorized and accessible. 

\begin{figure*}[t]
\centering
\includegraphics[width=1.95\columnwidth]{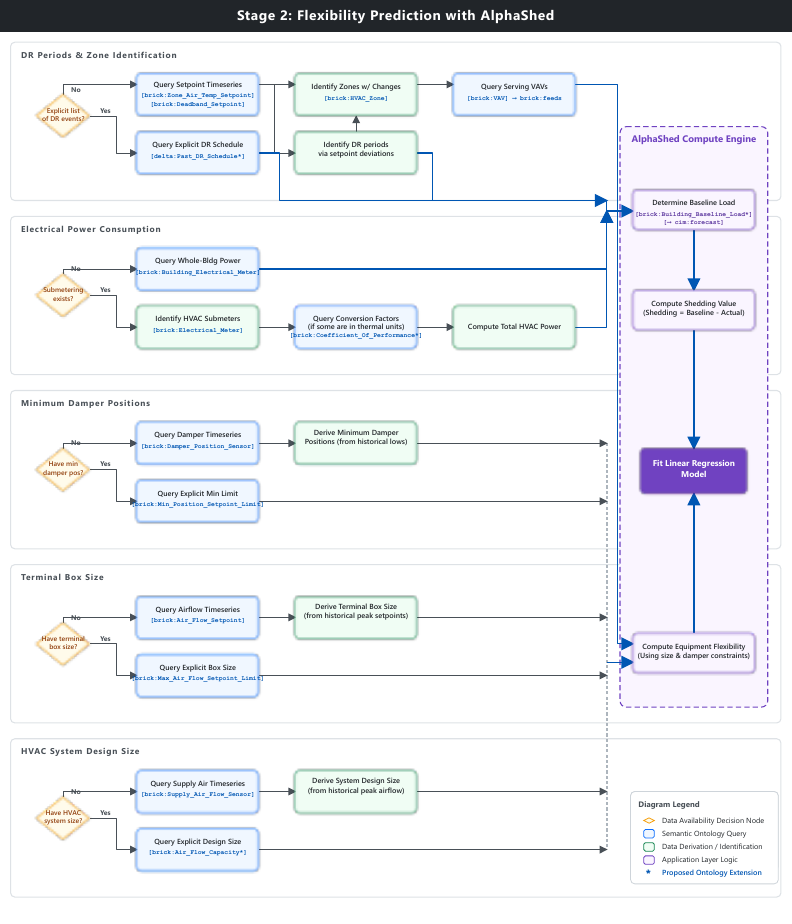}
\caption{A process flow diagram demonstrating the algorithmic implementation of AlphaShed \cite{vindel2023alphashed} using data and a Brick graph of the building.} 
\label{fig:alphashed23}
\end{figure*}

\subsection{Environmental Factors and Forecasts}

Unlike the building-specific energy baselines, IRs in this category can be modeled by the CIM standard. Through IEC 62325-301, CIM has classes for all identified real-time environmental factors including outside air temperature, relative humidity, wind characteristics, and direct/global solar irradiance—using the \texttt{AtmosphericAnalog} class paired with specific \texttt{kind} attributes (e.g., \texttt{ambientTemperature}, \texttt{irradianceGlobalHorizontal}).

Furthermore, CIM provides a highly effective and structured solution for the challenge that arises with forecasts. Rather than creating entirely new classes for predicted weather (e.g., creating a "PredictedTemperature" class), CIM utilizes a dedicated \texttt{Forecast} class. In this architecture, an \texttt{AtmosphericAnalog} entity is simply linked to a \texttt{Forecast} object, which defines the temporal prediction window and parameters. This is a major structural advantage, allowing the same physical measurement entities to be used for both real-time telemetry and future predictions.

While obtaining forecasts of weather-related events is easier by using the building's location, and this process can already be automated and saved in a database; Brick currently lacks this native separation of real-time versus predicted data. To clearly distinguish forecasts from real-time measurements in an open-source building ontology, we can define properties such as \texttt{hasForecastReference}, mimicking CIM's \texttt{Forecast} structure to separate prediction parameters from the standard taxonomy used for time series references.

\subsection{Proof of Concept: Implementation of AlphaShed}\label{sec:alphashed_proof}

To validate the proposed integrations, we implemented the AlphaShed algorithm \cite{vindel2023alphashed} utilizing real-world DR data and the corresponding Brick knowledge graph from FlexLAB\footnote{\url{https://flexlab.lbl.gov/}}. Figure \ref{fig:alphashed23} visualizes this implementation\footnote{Available at \url{https://github.com/ozanbarism/IRs4DR}}. AlphaShed was specifically selected for this application because the testbed is equipped with VAV dampers. This infrastructure provides the highly dynamic, time-series-intensive data necessary to demonstrate the proposed ontology extensions in practice.

During the implementation phase, several decision points emerged regarding data standardization, which are visualized in Figure \ref{fig:alphashed23}. For example, certain power measurements collected from the testbed were recorded in thermal units rather than electrical units. While load flexibility can be predicted at a macro, whole-building level, achieving granular predictions at the HVAC sub-system level requires standardizing these units. Consequently, it was necessary to introduce a COP variable to convert the thermal measurements into their electrical equivalents.

The accompanying process flow diagram illustrates this algorithmic workflow. It details the interplay between data availability, application logic, and the semantic queries made against the Brick graph. Additionally, the diagram highlights the current limitations of existing data models; the nodes marked with an asterisk (*) represent the specific ontology extensions required to successfully model these real-world DR concepts.

\end{document}